\documentclass[aps,prd,reprint,superscriptaddress,floatfix,nofootinbib,showkeys]{revtex4-2}
\usepackage{amsmath,amssymb}
\usepackage{graphicx}
\usepackage[colorlinks=true,
            linkcolor=blue,
            citecolor=blue,
            urlcolor=blue]{hyperref}
 \usepackage{orcidlink}
\begin{document}

\title{Phenomenology of Long-Lived Dark Photons and Axion-Like Particles in a Mixed Portal Framework}
\author{Caglar Zorbilmez\orcidlink{0000-0002-5199-061X}}
\email{czorbilmez@istanbul.edu.tr}
\author{Beyhan Tatar\orcidlink{0000-0003-3463-2009}}
\author{Azmi Ali Altintas\orcidlink{0000-0003-2383-4705}}
\affiliation{Department of Physics, Istanbul University, Faculty of Science, Istanbul, Turkey}
\author{Cihan Bayindir\orcidlink{0000-0002-3654-0469}}
\affiliation{Engineering Faculty, Istanbul Technical University, Istanbul, 34469 Turkey}
\author{Fatih Ozaydin\orcidlink{0000-0002-5872-9158}}
\affiliation{Data Science \& AI Major, Tokyo International University, 4-42-31 Higashi-Ikebukuro, Toshima-ku,
Tokyo 170-0013, Japan}
\affiliation{Institute for International Strategy and Emerging Technologies, Tokyo International University,
4-42-31 Higashi-Ikebukuro, Toshima-ku, Tokyo 170-0013, Japan}
\affiliation{Nanoelectronics Research Center, Istanbul, Türkiye}

\date{\today}

\begin{abstract}
We study a massive dark photon $A'$ and an axion-like particle $a$
connected through a gauge-invariant dimension-five portal to the
hypercharge field. After electroweak symmetry breaking, the portal
generates both the $a\gamma A'$ and the $aZA'$ couplings, with
$g_{aZ\gamma'}=-\tan\theta_W\,g_{a\gamma\gamma'}$, so the interaction that
allows $A'\to a\gamma$ also produces $Z\to aA'$. Together with
$a\to\gamma\gamma$, this gives the fully visible cascade
$Z\to a_1A'\to a_1a_2\gamma\to5\gamma$, in which the two ALPs are produced
at different points and can decay at macroscopic distances. We compute
the decay widths, including the full electroweak treatment of the
dark-photon width, the boosted decay lengths, and the sequential decay
probability, and compare the parameter space with existing constraints.
For $m_{A'}=10~\mathrm{GeV}$, $m_a=1~\mathrm{GeV}$, $\epsilon=10^{-7}$,
and $g_{a\gamma\gamma'}=g_{a\gamma\gamma}=10^{-6}~\mathrm{GeV}^{-1}$, we
find $\mathrm{Br}(Z\to aA')=2.92\times10^{-10}$, or about 440 produced
cascades per interaction point over the FCC-ee $Z$-pole program. We then
analyze the cascade with a fast simulation of the IDEA crystal
calorimeter that uses only photon energies and impact positions.
Displaced decays leave the momentum balance intact when all photons are
assumed to come from the interaction point, but they compress the
diphoton masses. A displaced-vertex fit that constrains each diphoton
pair to the ALP mass recovers the resonance structure of the cascade.
With a dedicated enriched simulation of the prompt $e^+e^-\to5\gamma$
background, the selection retains 576 signal events for four interaction
points, with a $95\%$ confidence-level upper limit of 2.8 background
events. This places the benchmark well above the $5\sigma$ discovery
threshold without any shower-direction information. In one-dimensional
coupling scans, the $5\sigma$ reach extends down to
$g_{a\gamma\gamma}\simeq2.5\times10^{-7}~\mathrm{GeV}^{-1}$ and
$g_{a\gamma\gamma'}\simeq1.5\times10^{-7}~\mathrm{GeV}^{-1}$; in
$g_{a\gamma\gamma}$ it has no upper edge within the scanned range and
extends into the prompt regime. Within the angular-response model used
here, a search based on non-pointing photons alone requires a core
shower-direction resolution of about $8~\mathrm{mrad}$ and a tail
probability $p_{\rm tail}\lesssim10^{-6}$ per photon to keep the
background below one event. With the displaced-vertex fit included, this
remains possible up to the largest core width scanned, $25.6~\mathrm{mrad}$,
for $p_{\rm tail}\lesssim10^{-3}$.

\end{abstract}


\maketitle

\section{Introduction}
\label{sec:introduction}

Long-lived particles (LLPs) provide well-motivated signatures of physics beyond the Standard Model. Their macroscopic lifetimes may arise from weak couplings, approximate symmetries, phase-space suppression, or interactions mediated by higher-dimensional operators. Depending on their decay lengths and production mechanisms, LLPs may generate displaced vertices, non-pointing photons, delayed energy deposits, disappearing tracks, or missing-energy signatures at collider experiments \cite{Alexander,Roeck,Hayrapetyan}. A broad experimental program has been developed to search for LLPs at present and future facilities, including dedicated concepts such as FASER, MATHUSLA, CODEX-b, ANUBIS, and proposed LLP-sensitive detector systems at future colliders \cite{LightLLP,Anubis,LLPatFCC,LLPatLHC,Mathusla,Faser,Codexb}.

Dark photons and axion-like particles (ALPs) are two widely studied examples of weakly coupled states. A dark photon, denoted by $A'$, may arise from an additional hidden Abelian gauge symmetry $U(1)_D$. Kinetic mixing between the hidden gauge field and the Standard Model hypercharge field induces a suppressed interaction between $A'$ and the electromagnetic current \cite{Holdom,Fabbrichesi}. The corresponding kinetic-mixing parameter $\epsilon$ controls both the visible production and the decay of the dark photon into electrically charged Standard Model particles. Searches at collider, fixed-target, beam-dump, and forward experiments have placed strong constraints on this interaction over a broad range of dark-photon masses and lifetimes \cite{BaBar2014,BaBar2017,FASER2024}.

Axion-like particles constitute another broad class of weakly interacting states predicted in many extensions of the Standard Model \cite{Irastorza,Jaeckel,Arvanitaki}. In contrast to the QCD axion, the mass and effective couplings of a generic ALP can be treated as independent phenomenological parameters. The interaction between an ALP and two photons permits the decay $a\to\gamma\gamma$, whose rate scales as $g_{a\gamma\gamma}^{2}m_a^3$. Consequently, ALPs may decay promptly, produce displaced diphoton vertices, or escape the detector, depending on their masses, couplings, and Lorentz boosts \cite{Bauer,Mimasu}.

Complementary collider studies have investigated direct ALP--fermion interactions, including electrophilic ALPs at the Electron--Ion Collider and Yukawa-like fermion-coupled ALPs produced in association with a photon at the LHC \cite{AdhikaryEIC,AdhikaryLHC}. These scenarios rely on direct ALP--fermion couplings and are complementary to, but distinct from, the photonic and mixed-portal framework considered in the present work.

A particularly interesting possibility arises when the dark-photon and ALP sectors are connected through the dark axion portal \cite{Kaneta,deNiverville}. The corresponding mixed interaction permits the exotic decay $A'\to a\gamma$ in addition to the conventional visible channel $A'\to f\overline{f}$. If the ALP subsequently decays into two photons, the dark-photon decay produces the three-photon cascade $A'\to a\gamma\to3\gamma$. The competition between the conventional and exotic dark-photon decay modes can therefore modify both the branching fractions and the expected LLP signatures.

A recent study investigated the axion--photon--dark-photon interaction at future lepton colliders, focusing primarily on single-photon plus missing-energy signatures in the regime where the ALP is sufficiently light and weakly coupled to escape the detector \cite{ChenHsiehTran2026}. The present work probes a complementary region of the same mixed-portal framework. We instead consider a massive ALP that decays visibly through $a\to\gamma\gamma$ and exploit resonant production at the FCC-ee (Z) pole through $Z\to aA'$. The subsequent decay $A'\to a\gamma$ generates the fully visible five-photon cascade $Z\to a_{1}A'\to a_{1}a_{2}\gamma\to5\gamma$, with the two diphoton decays occurring promptly or at displaced vertices. Our analysis complements the missing-energy search by addressing visible and long-lived ALP signatures and by incorporating the sequential decay geometry of the complete cascade.

The three-photon final state of $Z\to\gamma a$, $a\to\gamma\gamma$, has been studied for the FCC-ee with a parametrized simulation of the IDEA detector~\cite{Polesello2025}. That study does not use the direction of the calorimeter showers, because the pointing performance of the FCC-ee calorimeters has not yet been established. Our main selection follows the same principle.

Electroweak gauge invariance has an important consequence for this framework, as pointed out by Jod{\l}owski~\cite{Jodlowski2025}. When the dimension-five mixed portal is formulated in terms of the Standard Model hypercharge field before electroweak symmetry breaking, it generates both the $a\gamma A'$ and $aZA'$ interactions in the physical basis. Within this mixed-operator basis, their couplings satisfy
$g_{aZ\gamma'}=-\tan\theta_W\,g_{a\gamma\gamma'}$.
This relation follows from the gauge structure of the mixed operator itself and is independent of whether an additional ordinary ALP operator $aW^i_{\mu\nu}\widetilde W^{i\mu\nu}$ is included. The interaction responsible for $A'\to a\gamma$ therefore also permits the decay $Z\to aA'$. This provides a resonant and theoretically correlated production mechanism for the dark photon and the ALP at the FCC-ee $Z$ pole. Reference~\cite{Jodlowski2025} studied this production channel at LEP, the FCC-ee, and forward LHC detectors, with semi-visible displaced decays of the heavier dark state ($\gamma$ plus invisible, or $f^+f^-$ plus invisible) and missing-energy signatures. In those signatures the lighter dark state escapes detection, whereas here both ALPs decay to photons inside the detector and the cascade can be reconstructed entirely from photons.

The resulting exotic cascade is
$Z\to a_1A'\to a_1a_2\gamma\to5\gamma$,
where $a_1$ and $a_2$ denote the same ALP species but originate from different stages of the decay chain. The primary ALP $a_1$ is produced directly at the interaction point, whereas $a_2$ is produced at the dark-photon decay vertex. Their laboratory-frame boosts and decay lengths are generally different, and the probability that the complete cascade occurs inside the detector cannot be obtained by treating the dark photon and the two ALPs as particles produced from a common vertex with a fixed boost. A sequential treatment of the decay positions and exact two-body kinematics is required.

The high-statistics $Z$-pole program of the FCC-ee provides a particularly suitable environment for this search. The baseline FCC-ee program is expected to produce approximately $6\times10^{12}$ $Z$ bosons in total across four interaction points~\cite{FCC2025}. Assuming an equal luminosity share among the four interaction points, this corresponds to approximately $1.5\times10^{12}$ $Z$ bosons for a single detector. We use this number for the analytic estimates. The detector-level analysis follows the planned three-energy program, with $40$, $125$, and $40~\mathrm{ab}^{-1}$ at $\sqrt{s}=87.9$, $91.2$, and $94.3~\mathrm{GeV}$~\cite{FCC2025,Polesello2025}.
Depending on the lifetimes, the cascade may appear as a fully reconstructed five-photon event, contain one or two displaced diphoton vertices, or become partially invisible if an ALP decays outside the detector. Detector concepts proposed for the FCC-ee, including IDEA and complementary LLP-sensitive detector designs, provide additional motivation for studying such signatures \cite{IDEADetector,IDEA,Hecate}.

The phenomenology depends on the masses $m_{A'}$ and $m_a$, the kinetic-mixing parameter $\epsilon$, and the effective couplings $g_{a\gamma\gamma'}$ and $g_{a\gamma\gamma}$. These parameters cannot be selected independently of existing experimental and observational limits. In particular, the hypercharge completion also induces interactions involving the $Z$ boson, so constraints from processes such as $Z\to a\gamma$ and $Z\to aA'$ must be considered together with direct ALP searches, dark-photon searches, beam-dump limits, and astrophysical and cosmological bounds. For this reason, we first identify the relevant excluded regions and only then select representative benchmark points for the FCC-ee analysis.

In this work, we construct the gauge-invariant mixed-portal framework
and derive its interactions after electroweak symmetry breaking. We
calculate the partial decay widths for $Z\to aA'$,
$A'\to f\overline{f}$, $A'\to a\gamma$, and
$a\to\gamma\gamma$, together with the corresponding branching
fractions, exact two-body kinematics, and Lorentz-boosted decay lengths.
The competition between the Standard Model and exotic dark-photon decay
modes is characterized by the diagnostic ratio
$D=\Gamma(A'\to a\gamma)/\Gamma_{\mathrm{SM}}(A')$.
We emphasize that the condition $D=1$ identifies equal partial widths
but is not used by itself to define phenomenologically viable benchmark
parameters.

We then combine the production branching fraction, the dark-photon
branching ratio, and the sequential decay geometry to obtain the
five-photon fiducial yield. For the representative benchmark
$m_{A'}=10~\mathrm{GeV}$,
$m_a=1~\mathrm{GeV}$,
$\epsilon=10^{-7}$,
$g_{a\gamma\gamma'}=10^{-6}~\mathrm{GeV}^{-1}$, and
$g_{a\gamma\gamma}=10^{-6}~\mathrm{GeV}^{-1}$,
approximately $438$ cascades are produced in a single FCC-ee detector.
The spherical $R=2~\mathrm{m}$ reference calculation gives
$P_{\mathrm{seq}}=0.586$ and
$N_{5\gamma}^{\mathrm{fid}}\simeq257$.

We then go beyond this fiducial estimate with a fast detector-level
analysis based on the crystal calorimeter of the IDEA concept. The study
includes the displaced decay positions event by event, the calorimeter
geometry and resolution, photon-energy thresholds, cluster separation,
and the prompt Standard Model $e^+e^-\to5\gamma$ background. The main
selection uses only photon energies and impact positions. We show that
the displacement does not affect the momentum balance of the event when
all photons are assumed to come from the interaction point, but it
compresses the diphoton masses, and we introduce a displaced-vertex fit that constrains each diphoton pair to the ALP mass and restores the resonance structure of the cascade. The background after this selection is estimated with a
dedicated enriched simulation. For four interaction points the
selection retains 576 signal events, with a $95\%$ confidence-level (CL) upper limit of 2.8 background events. We then map the sensitivity as a function of the
couplings, discuss what non-pointing photons would add, and show how the
sensitivity depends on the masses.

The paper is organized as follows. In Sec.~II, we introduce the
gauge-invariant mixed-portal framework and derive the effective
interactions after electroweak symmetry breaking. In Sec.~III, we
calculate the relevant decay widths, branching fractions, exact
two-body kinematics, and laboratory-frame decay lengths. In Sec.~IV, we
discuss the existing experimental, astrophysical, and cosmological
constraints and define the benchmark-selection strategy. In Sec.~V, we
first present the analytic and fiducial FCC-ee phenomenology and then
develop the detector-level analysis: calorimeter resolvability, the displaced-vertex selection and its background, non-pointing photons, coupling sensitivity, and mass dependence. Finally, Sec.~VI summarizes our conclusions and
the limitations of the analysis.

\section{Gauge-Invariant Mixed-Portal Framework}
\label{sec:mixed-portal-framework}

\subsection{Field Content and Portal Interactions}

We extend the Standard Model (SM) by introducing a pseudoscalar
axion-like particle (ALP), denoted by $a$, and a massive dark photon,
denoted by $A'_{\mu}$, associated with a hidden $U(1)_{D}$ gauge
symmetry. The corresponding dark-photon field-strength tensor is
defined as
\begin{equation}
F_{D}^{\mu\nu}
=
\partial^{\mu}A'^{\nu}
-
\partial^{\nu}A'^{\mu}.
\label{eq:dark-photon-field-strength}
\end{equation}
The dark-photon mass is assumed to be generated in the hidden sector,
either through a dark-Higgs or a Stueckelberg mechanism. The details of
the mass-generation mechanism are not relevant for the phenomenological
analysis presented below.

For any field-strength tensor $X_{\mu\nu}$, we define the dual tensor as
\begin{equation}
\widetilde{X}^{\mu\nu}
=
\frac{1}{2}\varepsilon^{\mu\nu\rho\sigma}X_{\rho\sigma},
\qquad
\varepsilon^{0123}=+1.
\label{eq:dual-field-strength}
\end{equation}
This equation fixes both the sign and the normalization convention used
throughout the analysis.

Before electroweak symmetry breaking, the gauge-invariant portal
interactions relevant to this study are written as
\begin{equation}
\begin{aligned}
\mathcal{L}_{\mathrm{portal}}
={}&
-\frac{\epsilon_{Y}}{2}
B_{\mu\nu}F_{D}^{\mu\nu}
-\frac{C_{BD}}{2\Lambda}
a B_{\mu\nu}\widetilde{F}_{D}^{\mu\nu}
\\
&
-\frac{C_{BB}}{4\Lambda}
a B_{\mu\nu}\widetilde{B}^{\mu\nu},
\end{aligned}
\label{eq:gauge-invariant-portal}
\end{equation}
where $B_{\mu\nu}$ is the hypercharge field-strength tensor,
$\epsilon_{Y}$ is the hypercharge--dark-photon kinetic-mixing
parameter, $C_{BD}$ and $C_{BB}$ are dimensionless Wilson
coefficients, and $\Lambda$ is the effective scale of the theory.

The first term in Eq.~(\ref{eq:gauge-invariant-portal}) represents the
conventional vector portal. The second term is the mixed
ALP--hypercharge--dark-photon operator, whereas the third term is the
ordinary ALP--hypercharge operator. In the ordinary ALP sector,
electroweak gauge invariance also permits the independent dimension-five
operator
$a W^{i}_{\mu\nu}\widetilde{W}^{i\mu\nu}$.
For simplicity, we set its Wilson coefficient to zero and use
a minimal hypercharge completion for the ordinary ALP couplings. This
choice is not responsible for the mixed-portal relation derived below:
the operator $a W^{i}_{\mu\nu}\widetilde{W}^{i\mu\nu}$ contains no
dark-photon field strength and hence cannot modify
$g_{a\gamma\gamma'}$ or $g_{aZ\gamma'}$.

At dimension five, for an electroweak-singlet ALP and a hidden
$U(1)_D$ field strength, the mixed operator
$a B_{\mu\nu}\widetilde{F}_{D}^{\mu\nu}$ is gauge invariant,
whereas a term of the form
$a W^{i}_{\mu\nu}\widetilde{F}_{D}^{\mu\nu}$ is not an
$SU(2)_L$ singlet because the adjoint index $i$ is uncontracted.
Additional gauge-invariant mixed structures involving
$W^{i}_{\mu\nu}$ require extra electroweak insertions, for example
$H^\dagger \tau^i H$, and therefore arise at higher EFT dimension.
Such operators could relax the fixed mixed-coupling relation after
electroweak symmetry breaking, but they are outside the operator basis
considered in the present work.

\subsection{Effective Interactions after Electroweak Symmetry Breaking}

After electroweak symmetry breaking, the hypercharge field-strength
tensor can be expressed as
\begin{equation}
B_{\mu\nu}
=
\cos\theta_{W} F_{\mu\nu}
-
\sin\theta_{W} Z_{\mu\nu},
\label{eq:hypercharge-decomposition}
\end{equation}
where $F_{\mu\nu}$ and $Z_{\mu\nu}$ are the photon and $Z$-boson
field-strength tensors, respectively, and $\theta_{W}$ is the weak
mixing angle.

Keeping the interactions relevant to the production and decay channels
considered in this work, the effective interaction Lagrangian becomes
\begin{equation}
\begin{aligned}
\mathcal{L}_{\mathrm{int}}
={}&
\epsilon e A'_{\mu}J_{\mathrm{EM}}^{\mu}
-\frac{1}{2}g_{a\gamma\gamma'}
a F_{\mu\nu}\widetilde{F}_{D}^{\mu\nu}
\\
&
-\frac{1}{2}g_{aZ\gamma'}
a Z_{\mu\nu}\widetilde{F}_{D}^{\mu\nu}
-\frac{1}{4}g_{a\gamma\gamma}
a F_{\mu\nu}\widetilde{F}^{\mu\nu}
\\
&
-\frac{1}{2}g_{aZ\gamma}
a F_{\mu\nu}\widetilde{Z}^{\mu\nu}
-\frac{1}{4}g_{aZZ}
a Z_{\mu\nu}\widetilde{Z}^{\mu\nu},
\end{aligned}
\label{eq:effective-interaction-lagrangian}
\end{equation}
where
\begin{equation}
J_{\mathrm{EM}}^{\mu}
=
\sum_{f}Q_{f}\overline{f}\gamma^{\mu}f
\label{eq:electromagnetic-current}
\end{equation}
is the electromagnetic current, $Q_f$ is the electric charge of the
fermion $f$ in units of $e$, and
$\epsilon\simeq\epsilon_{Y}\cos\theta_{W}$ is the effective
photon--dark-photon kinetic-mixing parameter at leading order in the
small mixing. The coupling $\epsilon eA'_\mu J^\mu_{\mathrm{EM}}$ is the
leading term for $m_{A'}^2\ll m_Z^2$. The admixture of the $Z$ boson in
the dark photon, which matters at larger masses, is included in the
widths of Sec.~\ref{subsec:dark-photon-widths}.

The mixed portal couplings are
\begin{equation}
\begin{aligned}
g_{a\gamma\gamma'}
&=
\frac{C_{BD}}{\Lambda}\cos\theta_{W},
\\
g_{aZ\gamma'}
&=
-\frac{C_{BD}}{\Lambda}\sin\theta_{W}
=
-\tan\theta_{W}\,g_{a\gamma\gamma'}.
\end{aligned}
\label{eq:mixed-coupling-relations}
\end{equation}
The relation $g_{aZ\gamma'}=-\tan\theta_W\,g_{a\gamma\gamma'}$ was
derived in Ref.~\cite{Jodlowski2025}; we repeat the derivation to fix
the sign and normalization conventions used below.
Equation~(\ref{eq:mixed-coupling-relations}) is particularly important
for the present analysis because it shows that the dimension-five
hypercharge mixed portal necessarily generates the decay $Z\to aA'$.
The relation in Eq.~(\ref{eq:mixed-coupling-relations}) follows from
the mixed operator $aB_{\mu\nu}\widetilde F_D^{\mu\nu}$ and does
not rely on setting the ordinary
$aW^i_{\mu\nu}\widetilde W^{i\mu\nu}$ coefficient to zero.
Higher-dimensional mixed operators with additional electroweak
insertions could modify this relation and would introduce additional
EFT coefficients; such extensions are not included here.

In the ordinary ALP sector, the minimal hypercharge completion gives
\begin{equation}
\begin{aligned}
g_{a\gamma\gamma}
&=
\frac{C_{BB}}{\Lambda}\cos^{2}\theta_{W},
\\
g_{aZ\gamma}
&=
-\frac{C_{BB}}{\Lambda}\sin\theta_{W}\cos\theta_{W}
=
-\tan\theta_{W}\,g_{a\gamma\gamma},
\\
g_{aZZ}
&=
\frac{C_{BB}}{\Lambda}\sin^{2}\theta_{W}
=
\tan^{2}\theta_{W}\,g_{a\gamma\gamma}.
\end{aligned}
\label{eq:ordinary-alp-coupling-relations}
\end{equation}
For the mass hierarchy considered in this work,
$m_a<m_{A'}$ and $m_a+m_{A'}<m_Z$, the on-shell ALP decays involving
a $Z$ boson or a dark photon are kinematically forbidden. The dominant
ALP decay is therefore taken to be $a\to\gamma\gamma$. Nevertheless,
the coupling $g_{aZ\gamma}$ can induce $Z\to a\gamma$ and must be
included when existing experimental constraints are applied.

\subsection{Production and Decay Chains}

At the $Z$-pole stage of the FCC-ee, the interaction proportional to
$g_{aZ\gamma'}$ allows the resonant production process
\begin{equation}
e^{+}e^{-}
\to
Z
\to
a_{1}A',
\label{eq:fcc-production-process}
\end{equation}
provided that
\begin{equation}
m_a+m_{A'}<m_Z.
\label{eq:z-decay-kinematic-condition}
\end{equation}
The dark photon can subsequently decay either through the
kinetic-mixing portal,
\begin{equation}
A'\to f\overline{f},
\qquad
m_{A'}>2m_f,
\label{eq:dark-photon-fermionic-decay}
\end{equation}
or through the mixed portal,
\begin{equation}
A'\to a_{2}\gamma,
\qquad
m_{A'}>m_a.
\label{eq:dark-photon-alp-decay}
\end{equation}
Both ALPs may decay into photon pairs,
\begin{equation}
a_{1,2}\to\gamma\gamma.
\label{eq:alp-photon-decay}
\end{equation}
Here, $a_1$ and $a_2$ denote the same ALP species and are distinguished
only by their production origin: $a_1$ is produced directly in the
$Z$-boson decay, whereas $a_2$ originates from the subsequent
dark-photon decay.

The complete visible fermionic chain is therefore
\begin{equation}
Z
\to
a_1A'
\to
a_1 f\overline{f}
\to
\gamma\gamma f\overline{f},
\label{eq:fermionic-visible-chain}
\end{equation}
whereas the exotic cascade gives
\begin{equation}
Z
\to
a_1A'
\to
a_1a_2\gamma
\to
5\gamma.
\label{eq:five-photon-chain}
\end{equation}

In the numerical analysis, we impose the mass hierarchy
\begin{equation}
m_a<m_{A'},
\qquad
m_a+m_{A'}<m_Z.
\label{eq:mass-hierarchy}
\end{equation}
No numerical coupling ranges are fixed at this stage. The benchmark
values of $\epsilon$, $g_{a\gamma\gamma'}$, and
$g_{a\gamma\gamma}$ are selected only after the relevant collider,
beam-dump, astrophysical, and cosmological constraints have been
imposed.

\section{Production Rates, Decay Widths, and Lifetimes}
\label{sec:production-decays}

We derive the production and decay rates required for the event-yield and
decay-length analyses. In contrast to an approach based on an assumed
fixed Lorentz boost, the boosts of the particles produced at the $Z$ pole are determined directly from the corresponding two-body
kinematics.

\subsection{Production through \texorpdfstring{$Z\to aA'$}{Z to a A'}}
\label{subsec:z-production}

The interaction responsible for the production of an ALP and a dark
photon in an on-shell $Z$-boson decay is obtained from
Eq.~(\ref{eq:effective-interaction-lagrangian}):
\begin{equation}
\mathcal{L}_{aZ\gamma'}
=
-\frac{1}{2}g_{aZ\gamma'}
a Z_{\mu\nu}\widetilde{F}_{D}^{\mu\nu}.
\label{eq:a-z-dark-photon-interaction}
\end{equation}

It is convenient to introduce the K\"all\'en function
\begin{equation}
\lambda(x,y,z)
=
x^{2}+y^{2}+z^{2}-2xy-2xz-2yz.
\label{eq:kallen-function}
\end{equation}
The magnitude of the three-momentum of either final-state particle in
the $Z$-boson rest frame is then
\begin{equation}
p_{*}
=
\frac{\lambda^{1/2}\!\left(m_Z^{2},m_a^{2},m_{A'}^{2}\right)}
{2m_Z}.
\label{eq:z-decay-rest-frame-momentum}
\end{equation}

After summing over the final-state dark-photon polarizations and
averaging over the three polarization states of the initial $Z$ boson,
the squared matrix element is
\begin{equation}
\overline{\left|\mathcal{M}\right|^{2}}
=
\frac{g_{aZ\gamma'}^{2}}{6}
\lambda\!\left(m_Z^{2},m_a^{2},m_{A'}^{2}\right).
\label{eq:z-decay-squared-amplitude}
\end{equation}
Using the two-body phase-space formula, the corresponding partial decay
width is~\cite{Jodlowski2025}
\begin{equation}
\Gamma\!\left(Z\to aA'\right)
=
\frac{g_{aZ\gamma'}^{2}}{96\pi m_Z^{3}}
\left[
\lambda\!\left(m_Z^{2},m_a^{2},m_{A'}^{2}\right)
\right]^{3/2}.
\label{eq:z-to-a-dark-photon-width}
\end{equation}
This decay is kinematically allowed only when
$m_a+m_{A'}<m_Z$, as stated in
Eq.~(\ref{eq:z-decay-kinematic-condition}). The phase-space factor in
Eq.~(\ref{eq:z-to-a-dark-photon-width}) ensures that the partial width
vanishes at the kinematic threshold.

In the minimal hypercharge completion, the coupling relation in
Eq.~(\ref{eq:mixed-coupling-relations}) gives
\begin{equation}
\Gamma\!\left(Z\to aA'\right)
=
\frac{\tan^{2}\theta_W\,g_{a\gamma\gamma'}^{2}}
{96\pi m_Z^{3}}
\left[
\lambda\!\left(m_Z^{2},m_a^{2},m_{A'}^{2}\right)
\right]^{3/2}.
\label{eq:z-to-a-dark-photon-width-minimal}
\end{equation}
Therefore, the same mixed coupling $g_{a\gamma\gamma'}$ that controls
the decay $A'\to a\gamma$ also determines the production rate through
$Z\to aA'$.

The branching ratio of the $Z$ boson into the mixed-portal final state
is
\begin{equation}
\operatorname{Br}\!\left(Z\to aA'\right)
=
\frac{\Gamma\!\left(Z\to aA'\right)}
{\Gamma_Z^{\mathrm{tot}}},
\label{eq:z-to-a-dark-photon-branching-ratio}
\end{equation}
where $\Gamma_Z^{\mathrm{tot}}$ is the total width of the $Z$ boson.
For a sample containing $N_Z$ on-shell $Z$ bosons, the number of
produced $aA'$ pairs before detector acceptance and decay-probability
requirements is
\begin{equation}
N_{\mathrm{prod}}
=
N_Z\operatorname{Br}\!\left(Z\to aA'\right).
\label{eq:number-of-produced-a-dark-photon-pairs}
\end{equation}

At Born level, the nominal FCC-ee $Z$-pole calculation treats the
$Z$ boson as being at rest in the hard-process center-of-mass frame.
Initial-state radiation (ISR) and the beam-energy spread instead give an
event-dependent hard-process invariant mass and a small recoil of the
produced system.  The fixed-energy expressions below define the nominal
baseline; their impact on the benchmark boosts, decay lengths, and
detector acceptance is quantified by the dedicated stress test in
Sec.~\ref{subsubsec:idea-geometry}.  In this baseline, the energies of
the directly produced dark photon and ALP are
\begin{equation}
\begin{aligned}
E_{A'}
&=
\frac{m_Z^{2}+m_{A'}^{2}-m_a^{2}}{2m_Z},
\\
E_{a_1}
&=
\frac{m_Z^{2}+m_a^{2}-m_{A'}^{2}}{2m_Z}.
\end{aligned}
\label{eq:z-decay-final-state-energies}
\end{equation}
Their Lorentz boost factors are therefore fixed by the masses:
\begin{equation}
\begin{aligned}
(\beta\gamma)_{A'}
&=
\frac{p_*}{m_{A'}},
\\
(\beta\gamma)_{a_1}
&=
\frac{p_*}{m_a}.
\end{aligned}
\label{eq:z-decay-final-state-boosts}
\end{equation}
These mass-dependent boost factors are used in the decay-length and
fiducial decay-probability calculations instead of assigning a universal
benchmark value of $\beta\gamma$.

\subsection{Dark-Photon Decay Widths}
\label{subsec:dark-photon-widths}

The kinetic-mixing interaction in
Eq.~(\ref{eq:effective-interaction-lagrangian}) can be written as
\begin{equation}
\mathcal{L}_{A'f\bar f}
=
\epsilon e A'_{\mu}
\sum_f Q_f\overline{f}\gamma^{\mu}f.
\label{eq:dark-photon-fermion-interaction}
\end{equation}
For a charged lepton $\ell=e,\mu,\tau$, the corresponding partial
decay width is~\cite{Graham2021}
\begin{equation}
\Gamma\!\left(A'\to\ell^{+}\ell^{-}\right)
=
\frac{\alpha_{\mathrm{em}}\epsilon^{2}}{3}m_{A'}
\left(1+\frac{2m_{\ell}^{2}}{m_{A'}^{2}}\right)
\sqrt{1-\frac{4m_{\ell}^{2}}{m_{A'}^{2}}},
\label{eq:dark-photon-leptonic-width}
\end{equation}
where $\alpha_{\mathrm{em}}=e^{2}/(4\pi)$. Each channel contributes
only when $m_{A'}>2m_{\ell}$. The total leptonic contribution is
\begin{equation}
\Gamma_{\mathrm{lep}}(A')
=
\sum_{\ell=e,\mu,\tau}
\Gamma\!\left(A'\to\ell^{+}\ell^{-}\right)
\Theta\!\left(m_{A'}-2m_{\ell}\right),
\label{eq:dark-photon-total-leptonic-width}
\end{equation}
where $\Theta$ is the Heaviside step function.

In the low-mass region, a direct partonic calculation of the hadronic
width is not reliable because of non-perturbative QCD effects and
vector-meson resonances. We therefore introduce the experimentally
measured ratio~\cite{Navas2024}
\begin{equation}
\begin{aligned}
R_{\mathrm{had}}(s)
&=
\frac{\sigma\!\left(e^{+}e^{-}\to\mathrm{hadrons};s\right)}
{\sigma_{0}\!\left(e^{+}e^{-}\to\mu^{+}\mu^{-};s\right)},
\\
\sigma_{0}\!\left(e^{+}e^{-}\to\mu^{+}\mu^{-};s\right)
&=
\frac{4\pi\alpha_{\mathrm{em}}^{2}}{3s}.
\end{aligned}
\label{eq:r-had-definition}
\end{equation}
The inclusive hadronic width of the dark photon is then
\begin{equation}
\Gamma\!\left(A'\to\mathrm{hadrons}\right)
=
\frac{\alpha_{\mathrm{em}}\epsilon^{2}}{3}m_{A'}
R_{\mathrm{had}}\!\left(m_{A'}^{2}\right).
\label{eq:dark-photon-hadronic-width}
\end{equation}
In the numerical analysis, $R_{\mathrm{had}}(s)$ is evaluated
using experimental data in the resonance region and a perturbative
quark-level description only where it is reliable. This treatment
avoids introducing artificial discontinuities through a coarse
piecewise-constant approximation.

The total dark-photon width into Standard Model final states is
\begin{equation}
\Gamma_{\mathrm{SM}}(A')
=
\Gamma_{\mathrm{lep}}(A')
+
\Gamma\!\left(A'\to\mathrm{hadrons}\right).
\label{eq:dark-photon-sm-width}
\end{equation}

Equations~(\ref{eq:dark-photon-leptonic-width})--(\ref{eq:dark-photon-sm-width})
assume that the dark photon couples only to the electromagnetic current.
For hypercharge kinetic mixing this is the leading term in
$\delta\equiv m_{A'}^2/m_Z^2$. Diagonalizing the $(B,W^3,A')$ system to
first order in $\epsilon$, but to all orders in $\delta$, gives vector
and axial-vector couplings of the dark photon to a fermion
$f$~\cite{Curtin2015}
\begin{equation}
\begin{aligned}
C_V^f&=\epsilon e\left[Q_f+\frac{v_f}{\cos^2\theta_W}\,
\frac{\delta}{1-\delta}\right],\\
C_A^f&=\epsilon e\,\frac{a_f}{\cos^2\theta_W}\,
\frac{\delta}{1-\delta},
\end{aligned}
\label{eq:dark-photon-nc-couplings}
\end{equation}
with $v_f=T_3^f/2-\sin^2\theta_W Q_f$ and $a_f=T_3^f/2$. The
perturbative width into a fermion pair is then
\begin{equation}
\begin{aligned}
\Gamma\!\left(A'\to f\bar f\right)
={}&\frac{N_c\,m_{A'}}{12\pi}\sqrt{1-4x_f}\\
&\times\left[(C_V^f)^2(1+2x_f)+(C_A^f)^2(1-4x_f)\right],
\end{aligned}
\label{eq:dark-photon-nc-width}
\end{equation}
where $x_f=m_f^2/m_{A'}^2$ and $N_c$ is the number of colors. For
$\delta\to0$ this reduces to the electromagnetic result. The admixture
also opens the invisible channel
\begin{equation}
\Gamma\!\left(A'\to\nu\bar\nu\right)
=\frac{3\,m_{A'}}{12\pi}\,
2\left(\frac{\epsilon e}{4\cos^2\theta_W}\right)^{2}
\left(\frac{\delta}{1-\delta}\right)^{2},
\label{eq:dark-photon-nunu-width}
\end{equation}
summed over three neutrino flavors. In the numerical work we keep the
data-driven electromagnetic width of DarkCast v2.1~\cite{DarkCast} and
add to it the neutral-current terms, i.e., the difference between
Eq.~(\ref{eq:dark-photon-nc-width}) and its $\delta\to0$ limit, together
with Eq.~(\ref{eq:dark-photon-nunu-width}). These terms are evaluated
with $\alpha_{\mathrm{em}}=1/127.9$ and the fermion masses of the
MadGraph parameter card, the same inputs as in the event generation.
The expansion parameter satisfies $|\epsilon\tan\theta_W/(1-\delta)|<10^{-4}$ at every parameter point considered, so the first-order treatment is adequate over the scanned region. The correction to $\Gamma_{\mathrm{SM}}$ is $0.4\%$ at $m_{A'}=10~\mathrm{GeV}$ and increases rapidly toward the $Z$ pole: it is $5\%$ at $30~\mathrm{GeV}$, $24\%$ at $50~\mathrm{GeV}$, a factor $2.5$ at $70~\mathrm{GeV}$, and a factor $13.3$ at $m_{A'}=82.9~\mathrm{GeV}$, the kinematic endpoint of $Z\to aA'$ for $r=0.1$. At that endpoint neutrinos carry $16\%$ of
$\Gamma_{\mathrm{SM}}$.

The mixed portal provides the additional decay channel
$A'\to a\gamma$. The relevant interaction is
\begin{equation}
\mathcal{L}_{A'a\gamma}
=
-\frac{1}{2}g_{a\gamma\gamma'}
a F_{\mu\nu}\widetilde{F}_{D}^{\mu\nu},
\label{eq:dark-photon-alp-photon-interaction}
\end{equation}
which gives~\cite{Jodlowski2025}
\begin{equation}
\Gamma\!\left(A'\to a\gamma\right)
=
\frac{g_{a\gamma\gamma'}^{2}}{96\pi}m_{A'}^{3}
\left(1-\frac{m_a^{2}}{m_{A'}^{2}}\right)^{3}.
\label{eq:dark-photon-alp-photon-width}
\end{equation}
This channel is open for $m_{A'}>m_a$. Under the mass hierarchy
adopted in this work, the total dark-photon width is
\begin{equation}
\Gamma_{A'}^{\mathrm{tot}}
=
\Gamma_{\mathrm{SM}}(A')
+
\Gamma\!\left(A'\to a\gamma\right).
\label{eq:dark-photon-total-width}
\end{equation}

\subsection{ALP Decay Width}
\label{subsec:alp-width}

The ordinary ALP--photon interaction in
Eq.~(\ref{eq:effective-interaction-lagrangian}) is
\begin{equation}
\mathcal{L}_{a\gamma\gamma}
=
-\frac{1}{4}g_{a\gamma\gamma}
a F_{\mu\nu}\widetilde{F}^{\mu\nu}.
\label{eq:alp-photon-interaction}
\end{equation}
The corresponding diphoton decay width is~\cite{Mimasu}
\begin{equation}
\Gamma\!\left(a\to\gamma\gamma\right)
=
\frac{g_{a\gamma\gamma}^{2}}{64\pi}m_a^{3}.
\label{eq:alp-diphoton-width}
\end{equation}

The gauge-invariant completion also generates the interactions
$aZ\gamma$, $aZZ$, and $a\gamma A'$. However, the mass hierarchy
\begin{equation}
m_a<m_{A'},
\qquad
m_a+m_{A'}<m_Z
\label{eq:alp-decay-mass-hierarchy}
\end{equation}
implies that the on-shell decays $a\to\gamma A'$,
$a\to Z\gamma$, and $a\to ZZ$ are kinematically forbidden. Within
the minimal interaction set considered in this work, the total ALP width is
\begin{equation}
\Gamma_a^{\mathrm{tot}}
=
\Gamma\!\left(a\to\gamma\gamma\right),
\label{eq:alp-total-width}
\end{equation}
and consequently
\begin{equation}
\operatorname{Br}\!\left(a\to\gamma\gamma\right)=1.
\label{eq:alp-diphoton-branching-ratio}
\end{equation}

The labels $a_1$ and $a_2$ introduced in
Sec.~\ref{subsec:z-production} describe the same ALP species and hence
have the same proper decay width and proper lifetime. Their laboratory
decay lengths are nevertheless different because $a_1$ is produced
directly in the two-body decay $Z\to a_1A'$, whereas $a_2$ is produced
in the subsequent decay $A'\to a_2\gamma$ and has a different boost
distribution.

\subsection{Branching Ratios and Decay Lengths}
\label{subsec:branching-decay-lengths}

The proper lifetimes of the dark photon and the ALP are determined by
their total decay widths:
\begin{equation}
\tau_{A'}
=
\frac{\hbar}{\Gamma_{A'}^{\mathrm{tot}}},
\qquad
\tau_a
=
\frac{\hbar}{\Gamma_a^{\mathrm{tot}}}.
\label{eq:proper-lifetimes}
\end{equation}
Equivalently, their proper decay lengths are
\begin{equation}
c\tau_{A'}
=
\frac{\hbar c}{\Gamma_{A'}^{\mathrm{tot}}},
\qquad
c\tau_a
=
\frac{\hbar c}{\Gamma_a^{\mathrm{tot}}},
\label{eq:proper-decay-lengths}
\end{equation}
where~\cite{Navas2024}
\begin{equation}
\hbar c
=
1.97327\times10^{-16}\ \mathrm{GeV\,m}.
\label{eq:hbar-c-conversion}
\end{equation}

The mean decay length of a particle $i$ in the laboratory frame is
\begin{equation}
L_i
=
(\beta\gamma)_i c\tau_i
=
(\beta\gamma)_i
\frac{\hbar c}{\Gamma_i^{\mathrm{tot}}}.
\label{eq:general-laboratory-decay-length}
\end{equation}
For the dark photon and the directly produced ALP $a_1$, the boost
factors are fixed by the two-body $Z$-decay kinematics given in
Eq.~(\ref{eq:z-decay-final-state-boosts}). Their mean laboratory decay lengths are
\begin{equation}
\begin{aligned}
L_{A'}
&=
\frac{p_*}{m_{A'}}
\frac{\hbar c}{\Gamma_{A'}^{\mathrm{tot}}},
\\
L_{a_1}
&=
\frac{p_*}{m_a}
\frac{\hbar c}{\Gamma_a^{\mathrm{tot}}}.
\end{aligned}
\label{eq:dark-photon-direct-alp-decay-lengths}
\end{equation}

The boost of the secondary ALP $a_2$ is not described by a single
mass-dependent number in the $Z$-boson rest frame. In the dark-photon
rest frame, its energy and momentum are
\begin{equation}
E_{a_2}^{*}
=
\frac{m_{A'}^{2}+m_a^{2}}{2m_{A'}},
\qquad
p_{a_2}^{*}
=
\frac{m_{A'}^{2}-m_a^{2}}{2m_{A'}}.
\label{eq:secondary-alp-dark-photon-frame-kinematics}
\end{equation}
If $\theta^{*}$ denotes the angle between the $a_2$ momentum in the
dark-photon rest frame and the direction of motion of the dark photon,
then the $a_2$ energy in the $Z$-boson rest frame is
\begin{equation}
E_{a_2}(\theta^{*})
=
\gamma_{A'}
\left(
E_{a_2}^{*}
+
\beta_{A'}p_{a_2}^{*}\cos\theta^{*}
\right),
\label{eq:secondary-alp-laboratory-energy}
\end{equation}
where
\begin{equation}
\gamma_{A'}
=
\frac{E_{A'}}{m_{A'}},
\qquad
\beta_{A'}
=
\frac{p_*}{E_{A'}}.
\label{eq:dark-photon-beta-gamma}
\end{equation}
Consequently,
\begin{equation}
\begin{aligned}
(\beta\gamma)_{a_2}(\theta^{*})
&=
\frac{\sqrt{E_{a_2}^{2}(\theta^{*})-m_a^{2}}}{m_a},
\\
L_{a_2}(\theta^{*})
&=
(\beta\gamma)_{a_2}(\theta^{*})
\frac{\hbar c}{\Gamma_a^{\mathrm{tot}}}.
\end{aligned}
\label{eq:secondary-alp-boost-and-decay-length}
\end{equation}
The decay probability of $a_2$ must therefore be averaged over the
appropriate angular distribution rather than evaluated with a fixed
boost factor.

For a particle with mean laboratory decay length $L_i$, the probability
of decaying between path lengths $\ell_{\min}$ and $\ell_{\max}$ is
\begin{equation}
P_i\!\left(\ell_{\min},\ell_{\max}\right)
=
\exp\!\left(-\frac{\ell_{\min}}{L_i}\right)
-
\exp\!\left(-\frac{\ell_{\max}}{L_i}\right).
\label{eq:fiducial-decay-probability}
\end{equation}
In the detector analysis, the path-length boundaries are determined from the detector geometry and the particle direction. For the
sequential chain $A'\to a_2\gamma$, the displacement of the dark-photon
decay vertex must also be included when evaluating the fiducial decay
probability of $a_2$.

The branching fraction of the dark photon into the exotic channel is
\begin{equation}
\operatorname{Br}\!\left(A'\to a\gamma\right)
=
\frac{\Gamma\!\left(A'\to a\gamma\right)}
{\Gamma_{A'}^{\mathrm{tot}}}.
\label{eq:dark-photon-exotic-branching-ratio}
\end{equation}
It is useful to retain the dominance parameter
\begin{equation}
D
=
\frac{\Gamma\!\left(A'\to a\gamma\right)}
{\Gamma_{\mathrm{SM}}(A')},
\label{eq:dominance-parameter}
\end{equation}
in terms of which
\begin{equation}
\operatorname{Br}\!\left(A'\to a\gamma\right)
=
\frac{D}{1+D},
\qquad
\operatorname{Br}\!\left(A'\to\mathrm{SM}\right)
=
\frac{1}{1+D}.
\label{eq:branching-ratios-in-terms-of-d}
\end{equation}
Thus, $D=1$ corresponds to equal exotic and Standard Model decay
widths. The parameter $D$ is used only as a diagnostic quantity; it is
not used to select benchmark couplings. Benchmark points are chosen after applying the relevant experimental, astrophysical, and
cosmological constraints.

Since both ALPs decay into photon pairs in the minimal scenario, the
branching fraction for the complete five-photon cascade is
\begin{equation}
\begin{aligned}
&\operatorname{Br}\!\left(
Z\to a_1A'\to a_1a_2\gamma\to5\gamma
\right)
\\
&\qquad=
\operatorname{Br}\!\left(Z\to aA'\right)
\operatorname{Br}\!\left(A'\to a\gamma\right)
\left[
\operatorname{Br}\!\left(a\to\gamma\gamma\right)
\right]^{2}
\\
&\qquad=
\operatorname{Br}\!\left(Z\to aA'\right)
\operatorname{Br}\!\left(A'\to a\gamma\right).
\end{aligned}
\label{eq:five-photon-cascade-branching-ratio}
\end{equation}
Equation~(\ref{eq:five-photon-cascade-branching-ratio}) gives the
particle-level branching fraction. Detector acceptance, reconstruction
efficiencies, and the fiducial decay probabilities of $A'$, $a_1$, and
$a_2$ must be included separately when calculating the observable
five-photon event yield.
%%
%%%%

\section{Existing Constraints and Benchmark Selection}
\label{sec:constraints}

\subsection{Constraint Strategy and Parameter Slices}
\label{subsec:constraint-strategy}

The phenomenology of the mixed-portal framework depends on the
five-dimensional parameter set
\begin{equation}
\mathcal{P}
=
\left\{
m_{A'},\,
m_a,\,
\epsilon,\,
g_{a\gamma\gamma'},\,
g_{a\gamma\gamma}
\right\}.
\label{eq:full-parameter-set}
\end{equation}
Experimental limits cannot, in general, be interpreted as independent
one-dimensional bounds on these parameters. In particular, the exotic
decay $A'\to a\gamma$ changes both the visible branching fractions and
the lifetime of the dark photon. Consequently, limits derived for a
minimal dark photon that decays only into Standard Model particles
cannot be directly superimposed on the present parameter space without
accounting for these changes.

To display the multidimensional constraints in two-dimensional planes,
we define the mass ratio
\begin{equation}
r
=
\frac{m_a}{m_{A'}},
\qquad
0<r<1.
\label{eq:alp-dark-photon-mass-ratio}
\end{equation}
The main numerical scans are presented for the reference slice
\begin{equation}
r=0.1,
\label{eq:reference-mass-ratio}
\end{equation}
while selected alternative values are used to assess the dependence
on the mass hierarchy. For a fixed value of $r$, the condition for
on-shell production in $Z$-boson decay becomes
\begin{equation}
m_{A'}
<
\frac{m_Z}{1+r}.
\label{eq:dark-photon-mass-upper-bound}
\end{equation}
For the reference choice $r=0.1$, this gives
\begin{equation}
m_{A'}<82.9\ \mathrm{GeV}.
\label{eq:dark-photon-mass-upper-bound-reference}
\end{equation}
Thus, the $m_{A'}=100\ \mathrm{GeV}$ endpoint used in the preliminary
analysis is not part of the kinematically accessible $Z\to aA'$ region and is not retained.

We organize the existing constraints into three classes. First,
dark-photon searches constrain the kinetic-mixing parameter $\epsilon$
through production and decay into Standard Model final states. Second,
dark-axion-portal searches constrain $g_{a\gamma\gamma'}$ through
processes such as $e^{+}e^{-}\to aA'$ and $Z\to aA'$. Gauge invariance
requires the latter process through
$g_{aZ\gamma'}=-\tan\theta_W g_{a\gamma\gamma'}$~\cite{Jodlowski2025}, as discussed in
Sec.~\ref{sec:mixed-portal-framework}. Third, ALP searches constrain
$g_{a\gamma\gamma}$ through the production and decay of
$a\to\gamma\gamma$.

For a visible dark-photon search in a final state $X$, the expected
number of signal events has the schematic form
\begin{equation}
N_X
=
\mathcal{L}\,
\sigma_{\mathrm{prod}}\!\left(m_{A'},\epsilon\right)
\operatorname{Br}\!\left(A'\to X\right)
P_{\mathrm{fid}}\!\left(\tau_{A'}\right)
\mathcal{A}_X\varepsilon_X,
\label{eq:recast-event-yield}
\end{equation}
where $\mathcal{L}$ is the integrated luminosity,
$P_{\mathrm{fid}}$ is the relevant fiducial decay probability,
$\mathcal{A}_X$ is the geometrical and kinematic acceptance, and
$\varepsilon_X$ is the reconstruction and selection efficiency. The
production cross section in conventional kinetic-mixing searches is
unchanged by the additional decay channel, whereas the branching
fraction and lifetime appearing in
Eq.~(\ref{eq:recast-event-yield}) are modified. We therefore recast the
visible dark-photon limits following the event-yield procedure of
Ref.~\cite{Chen}, rather than treating the minimal-dark-photon
exclusion contours as model-independent bounds.

The mixed coupling $g_{a\gamma\gamma'}$ is tested independently using
existing dark-axion-portal searches, including constraints from
$Z$-boson factories, electron--positron colliders, beam-dump
experiments, and forward detectors. The fixed relation between
$g_{aZ\gamma'}$ and $g_{a\gamma\gamma'}$ is imposed when applying the
$Z$-boson constraints~\cite{Jodlowski2025}.

Constraints on $g_{a\gamma\gamma}$ are taken from direct ALP searches and from the constraint compilations of Refs.~\cite{OHare2024,AxionLimits}. We distinguish these limits according to their underlying physical assumptions. Direct laboratory bounds and astrophysical or cosmological constraints that do not require the ALP to constitute dark matter are imposed as exclusions. By contrast, indirect-detection bounds that assume a specified relic abundance, in particular that ALPs account for all of the galactic or cosmological dark matter, are not treated as model-independent exclusions in the present analysis.

\begin{figure}[t]
\centering
\includegraphics[width=0.98\linewidth]{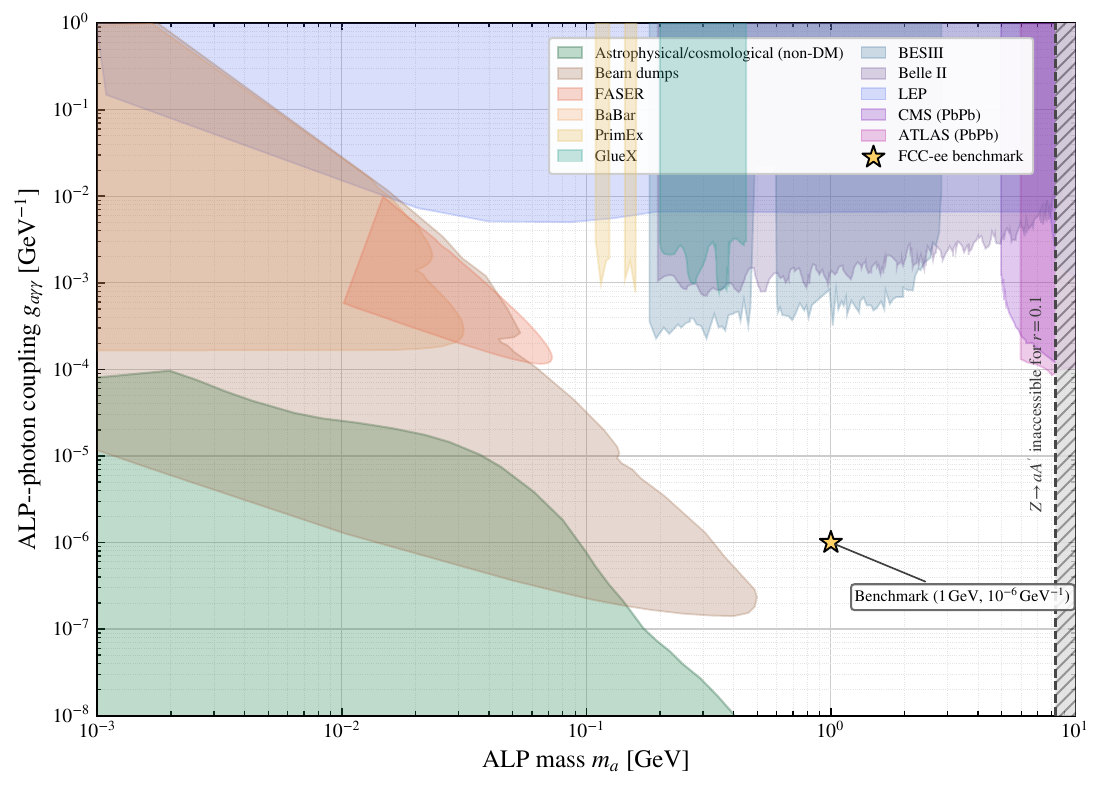}
\caption{Constraints on the ALP--photon coupling in the $(m_a,g_{a\gamma\gamma})$ plane. The individually labelled colored regions are excluded by direct laboratory searches. The green region represents the combined astrophysical and cosmological exclusions that do not assume that ALPs constitute dark matter, using the \texttt{Combined\_Astro} data set compiled in Ref.~\cite{AxionLimits}. Since $\mathrm{Br}(a\to\gamma\gamma)=1$ in the minimal scenario considered here, the direct diphoton limits can be applied without an additional branching-ratio rescaling. The star denotes the FCC-ee benchmark point, $(m_a,g_{a\gamma\gamma})=(1~\mathrm{GeV},10^{-6}~\mathrm{GeV}^{-1})$. The hatched region is kinematically inaccessible through $Z\to aA'$ for the reference hierarchy $r=m_a/m_{A'}=0.1$. Bounds that require ALPs to constitute a specified fraction of dark matter are neither displayed nor imposed.}
\label{fig:alp-photon-constraints}
\end{figure}

As shown in Fig.~\ref{fig:alp-photon-constraints}, the selected FCC-ee benchmark lies outside both the direct laboratory exclusions and the displayed dark-matter-independent astrophysical and cosmological exclusion region. For this benchmark, the diphoton width corresponds to a proper lifetime of approximately $\tau_a\simeq1.3\times10^{-10}~\mathrm{s}$. The ALP is therefore far too short-lived to constitute present-day dark matter. Consequently, limits based on the assumption of a galactic or cosmological ALP dark-matter population are not applicable to this point. The benchmark is therefore not excluded by the constraints displayed in Fig.~\ref{fig:alp-photon-constraints}; this is not the result of a complete global recast of every possible ALP search.
The high-mass edge of the dark-matter-independent astrophysical exclusion region, at ALP masses of order a few $10^{-1}~\mathrm{GeV}$, mainly reflects the weakening of supernova limits. Once the ALP mass exceeds several times the characteristic supernova-core temperature, $T_{\mathrm{SN}}\sim30~\mathrm{MeV}$, thermal ALP production becomes exponentially Boltzmann suppressed \cite{Giannotti2011,Caputo2022}. The precise endpoint is analysis dependent because the displayed region combines several astrophysical and cosmological constraints.

The published LEP constraints on the mixed portal were obtained in the
limit of a nearly massless ALP~\cite{Jodlowski2025}. To adapt these
contours to the reference slice $r=m_a/m_{A'}=0.1$, we apply the
leading two-body production phase-space rescaling

\begin{equation}
g_{a\gamma\gamma'}^{(r)}
=
g_{a\gamma\gamma'}^{(0)}
\left[
\frac{
\lambda\!\left(s,0,m_{A'}^{2}\right)
}{
\lambda\!\left(s,r^{2}m_{A'}^{2},m_{A'}^{2}\right)
}
\right]^{3/4},
\label{eq:mixed-portal-rescaling}
\end{equation}
where the K\"all\'en function is defined in
Eq.~(\ref{eq:kallen-function}). Here, $s=m_Z^2$ is used for the
$Z$-pole constraints, while $\sqrt{s}=205.4~\mathrm{GeV}$ is used for
the representative LEP2 constraint. This rescaling preserves the
leading production rate, which is proportional to
$g_{a\gamma\gamma'}^{2}\lambda^{3/2}$.

\begin{figure}[t]
    \centering
    \includegraphics[width=0.98\linewidth]
    {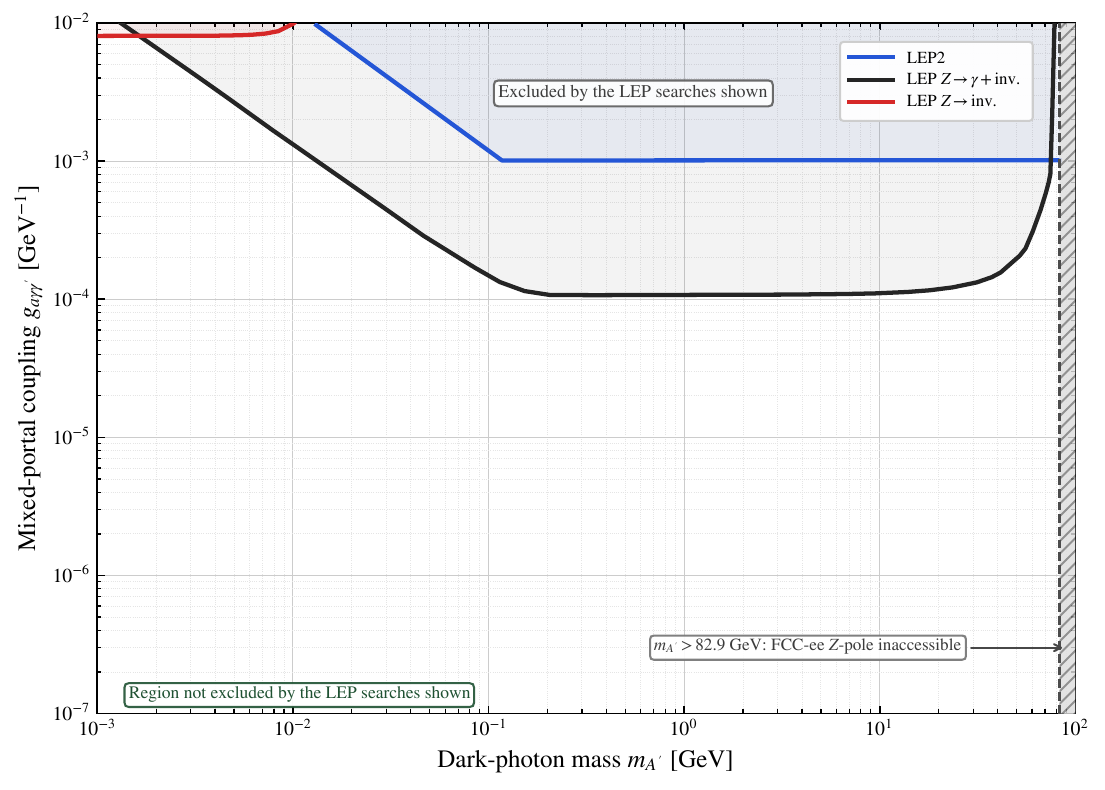}
    \caption{
    Existing LEP constraints in the
    $(m_{A'},g_{a\gamma\gamma'})$ plane.
    The published contours of Ref.~\cite{Jodlowski2025}, obtained for
    $m_a\simeq0$, have been rescaled to the reference hierarchy
    $r=m_a/m_{A'}=0.1$ using the leading production phase-space
    correction in Eq.~(\ref{eq:mixed-portal-rescaling}).
    The shaded regions are excluded by the indicated LEP searches.
    Acceptance and lifetime effects have not been re-simulated, so the
    rescaled contours are used as an approximate phenomenological
    preselection. The hatched region is inaccessible through
    on-shell $Z\to aA'$ production at FCC-ee for $r=0.1$.
    }
   \label{fig:mixed_portal_constraints}
\end{figure}

For $r=0.1$, the decay phase-space factor entering
$\Gamma(A'\to a\gamma)$ becomes
$(1-r^2)^3\simeq 0.9703$. Lifetime and acceptance effects are not
re-simulated in the LEP phase-space-only recast shown in
Fig.~\ref{fig:mixed_portal_constraints}; therefore, those contours are
interpreted as an approximate phenomenological preselection. In
contrast, the experiment-specific prompt-decay probabilities are
included in the dark-photon recast discussed below.

\subsection{Recasting of Prompt Visible Dark-Photon Constraints}
\label{subsec:visible_recast}

To quantify the effect of the exotic decay channel on the conventional dark-photon constraints, we recast the prompt visible-search limits implemented in DarkCast v2.1 \cite{DarkCast}. Let $\epsilon_{0,i}(m_{A'})$ denote the published kinetic-mixing limit obtained by the $i$th experiment in the minimal dark-photon scenario, for which $g_{a\gamma\gamma'}=0$. The Standard Model contribution to the total dark-photon decay width can be written as
\begin{equation}
\Gamma_{\mathrm{SM}}(A')=
\epsilon^{2}K(m_{A'}),
\label{eq:dark_photon_sm_width}
\end{equation}
where $K(m_{A'})$ contains the contributions of all kinematically accessible leptonic and hadronic final states.

After including the exotic decay channel $A'\to a\gamma$, the expected number of prompt visible events is rescaled according to
\begin{equation}
\mathcal{R}_{i}(\epsilon)=
\left(
\frac{\epsilon}{\epsilon_{0,i}}
\right)^{2}
\frac{
\epsilon^{2}K(m_{A'})
}{
\epsilon^{2}K(m_{A'})
+
\Gamma(A'\to a\gamma)
}
\frac{
P_i(\tau)
}{
P_i(\tau_{0,i})
}.
\label{eq:visible_recast_ratio}
\end{equation}
Here, $\tau$ is the dark-photon lifetime after including the exotic decay width, whereas $\tau_{0,i}$ is the lifetime corresponding to the published minimal-dark-photon limit $\epsilon_{0,i}$. The prompt-decay probability for the $i$th experiment is
\begin{equation}
P_i(\tau)=
1-
\exp\left(
-\frac{t_{\mathrm{max},i}}{\tau}
\right),
\label{eq:prompt_probability}
\end{equation}
where $t_{\mathrm{max},i}$ denotes the maximum proper decay time accepted by the corresponding experimental analysis.

\begin{figure}[htbp]
\centering
\includegraphics[width=0.98\linewidth]
{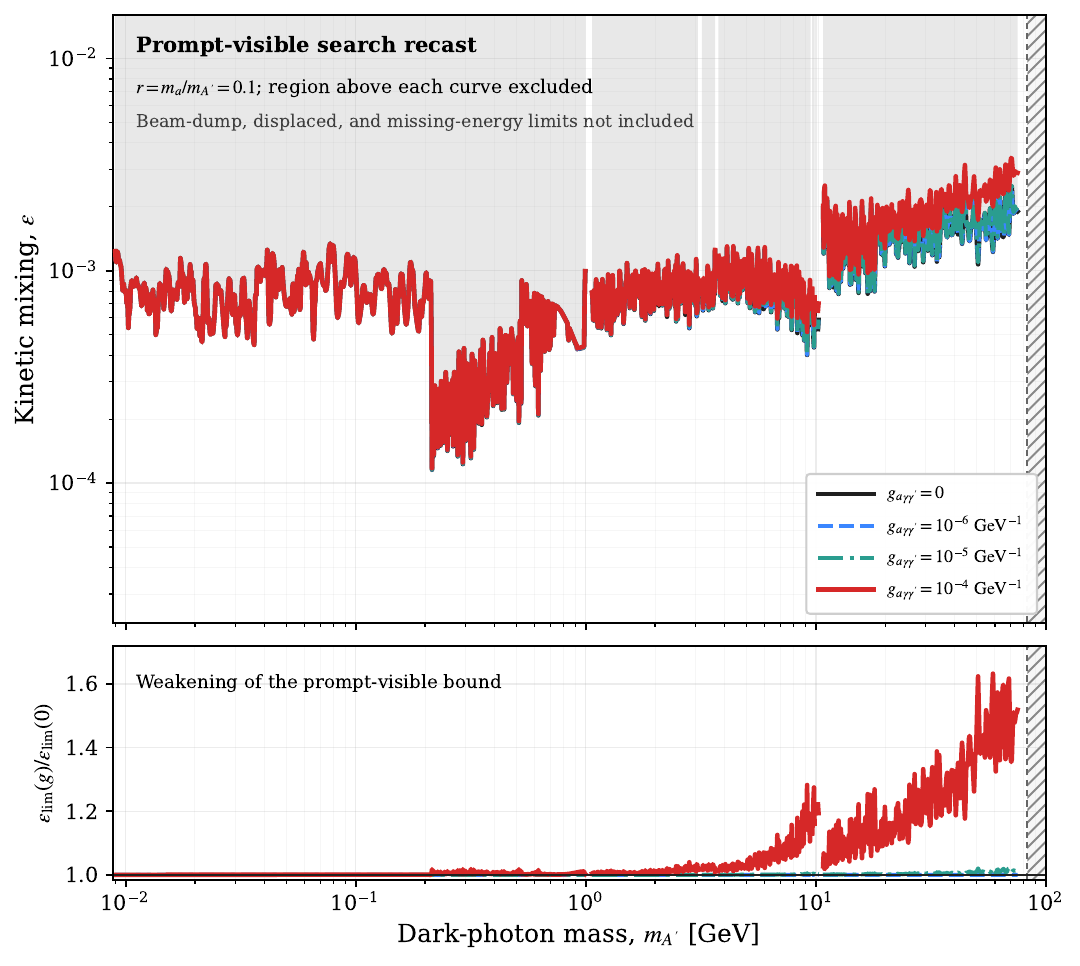}
\caption{Recast of the prompt visible dark-photon searches in the $(m_{A'},\epsilon)$ plane for the fixed mass ratio $r=m_a/m_{A'}=0.1$.  The upper panel shows the exclusion envelopes obtained for
$g_{a\gamma\gamma'}=0$,
$g_{a\gamma\gamma'}=10^{-6}\,\mathrm{GeV}^{-1}$,
$g_{a\gamma\gamma'}=10^{-5}\,\mathrm{GeV}^{-1}$, and
$g_{a\gamma\gamma'}=10^{-4}\,\mathrm{GeV}^{-1}$. For each fixed value of $g_{a\gamma\gamma'}$, the region above the corresponding curve is excluded by the prompt visible searches included in the recast. The gray-shaded region represents the conventional minimal-dark-photon exclusion obtained for $g_{a\gamma\gamma'}=0$. The lower panel shows the weakening factor $\epsilon_{\mathrm{lim}}(g_{a\gamma\gamma'})/\epsilon_{\mathrm{lim}}(0)$. The hatched region does not satisfy $m_a+m_{A'}<m_Z$ and is therefore inaccessible through the on-shell decay $Z\to aA'$ at the FCC-ee $Z$ pole. Beam-dump, displaced-decay, and missing-energy constraints are not included because their reinterpretation requires a dedicated treatment of the decay geometry and detector acceptance. The Standard Model width used here is the DarkCast v2.1 width without the neutral-current terms of Sec.~\ref{subsec:dark-photon-widths}; above about $50~\mathrm{GeV}$ these terms strengthen the prompt constraints (see text).}
\label{fig:dark_photon_visible_recast}
\end{figure}

For each experiment, the modified exclusion boundary is obtained by solving
\begin{equation}
\mathcal{R}_{i}(\epsilon)=1.
\label{eq:recast_condition}
\end{equation}
At each value of $m_{A'}$, the most restrictive result defines the prompt visible-search exclusion envelope shown in Fig.~\ref{fig:dark_photon_visible_recast}.

For $g_{a\gamma\gamma'}=10^{-6}\,\mathrm{GeV}^{-1}$, the modification
of the prompt visible-search limits is negligible over the considered
mass range. For $g_{a\gamma\gamma'}=10^{-5}\,\mathrm{GeV}^{-1}$, the
maximum weakening is approximately $2.1\%$. The effect becomes more
pronounced for $g_{a\gamma\gamma'}=10^{-4}\,\mathrm{GeV}^{-1}$, for
which the upper bound on $\epsilon$ can be weakened by a factor of
approximately $1.63$ in the high-mass region. Thus, the exotic decay
channel reduces the sensitivity of conventional visible dark-photon
searches but does not completely remove the corresponding experimental
constraints. The recast in Fig.~\ref{fig:dark_photon_visible_recast}
uses the DarkCast width without the neutral-current terms of
Eqs.~(\ref{eq:dark-photon-nc-couplings})--(\ref{eq:dark-photon-nunu-width}).
For a prompt limit the weakening factor is
$W^2=[1+\sqrt{1+4\Gamma(A'\to a\gamma)/(\epsilon_0^2K)}]/2$, where
$\Gamma_{\mathrm{SM}}=\epsilon^2K$. With the larger $K$ the high-mass
factor for $g_{a\gamma\gamma'}=10^{-4}~\mathrm{GeV}^{-1}$ becomes $1.56$,
$1.50$, $1.38$, $1.19$, and $1.12$ at $m_{A'}=50$, $60$, $70$, $80$, and
$82.9~\mathrm{GeV}$. Above about $50~\mathrm{GeV}$ the figure therefore
slightly understates the prompt constraints. This does not affect the
benchmark at $m_{A'}=10~\mathrm{GeV}$.

The parameter region retained for the numerical analysis is the part
of parameter space not excluded by the constraints included in this
work, i.e., the intersection
\begin{equation}
\mathcal{P}_{\mathrm{retained}}
=
\mathcal{P}_{\mathrm{kin}}
\cap
\mathcal{P}_{A'}
\cap
\mathcal{P}_{a\gamma\gamma'}
\cap
\mathcal{P}_{a\gamma\gamma},
\label{eq:allowed-parameter-intersection}
\end{equation}
where $\mathcal{P}_{\mathrm{kin}}$ denotes the kinematically allowed
region and the remaining sets denote the regions surviving the
dark-photon, mixed-portal, and ALP constraints, respectively. Excluded regions are displayed explicitly in every mass--coupling plot.
Benchmark points are selected only from
$\mathcal{P}_{\mathrm{retained}}$; no coupling interval is fixed before
this combined constraint analysis is performed.

\section{Numerical Analysis and FCC-ee Phenomenology}
\label{sec:numerical-analysis}

This section evaluates the FCC-ee phenomenology of the mixed dark-photon--ALP portal. We first examine the
competition between the Standard Model and exotic dark-photon decay
channels and then study the Lorentz-boosted decay lengths and fiducial
decay probabilities of the dark photon and the two ALPs appearing in
the cascade. These ingredients are combined to obtain the sequential
five-photon fiducial yield. We subsequently go beyond the spherical
fiducial estimate by performing an IDEA-oriented detector-level study
of the complete five-photon final state, including the decay geometry,
ECAL acceptance and cluster separation, photon-energy requirements, a displaced-vertex selection, non-pointing observables, Standard Model background rejection, and the
dependence of the selected signal yield on the masses and portal
couplings.

Throughout the analysis, the production process is taken to be
\begin{equation}
e^{+}e^{-}\to Z\to a_{1}A',
\end{equation}
at the FCC-ee $Z$ pole. The Lorentz boosts are calculated from the
two-body kinematics given in
Eqs.~(\ref{eq:z-decay-rest-frame-momentum})--%
(\ref{eq:z-decay-final-state-boosts}), rather than assigning a fixed
boost factor to the produced particles.

\subsection{Dark-Photon Branching Fractions and Channel Dominance}
\label{subsec:dark-photon-branching-fractions}

We begin by examining the competition between the conventional
Standard Model decay modes and the exotic channel
$A'\to a\gamma$. For the reference parameter slice, we set
\begin{equation}
\epsilon=10^{-7},
\qquad
r\equiv\frac{m_a}{m_{A'}}=0.1.
\label{eq:branching-ratio-reference-slice}
\end{equation}
The ALP mass therefore varies with the dark-photon mass according to
$m_a=0.1m_{A'}$. The kinematic condition for the production process
$Z\to aA'$ restricts the dark-photon mass to
\begin{equation}
0.1~{\rm GeV}
\leq m_{A'}
<
\frac{m_Z}{1+r}
\simeq 82.9~{\rm GeV}.
\label{eq:branching-ratio-mass-range}
\end{equation}

The following representative values of the mixed-portal coupling are
considered:
\begin{equation}
g_{a\gamma\gamma'}
=
3\times10^{-9},\;
10^{-8},\;
3\times10^{-8},\;
10^{-7},\;
3\times10^{-7}
\ {\rm GeV}^{-1}.
\label{eq:mixed-portal-branching-benchmarks}
\end{equation}
These values provide a constraint-informed phenomenological scan of the
transition between the Standard Model and cascade-dominated regimes.
They should not be interpreted as implying that every mass point along
each curve is free of experimental constraints. Only the portions satisfying the
combined condition defined in
Eq.~(\ref{eq:allowed-parameter-intersection}) are retained as viable
parameter points.

The Standard Model contribution $\Gamma_{\rm SM}(A')$ is evaluated
using the leptonic and mass-dependent hadronic widths implemented in
DarkCast v2.1, to which we add the neutral-current and $A'\to\nu\bar\nu$
terms of Sec.~\ref{subsec:dark-photon-widths}. The exotic width is
calculated from Eq.~(\ref{eq:dark-photon-alp-photon-width}). The
resulting exotic branching fraction is
\begin{equation}
\operatorname{Br}(A'\to a\gamma)
=
\frac{\Gamma(A'\to a\gamma)}
{\Gamma_{\rm SM}(A')+\Gamma(A'\to a\gamma)}
=
\frac{D}{1+D},
\label{eq:numerical-exotic-branching-ratio}
\end{equation}
where $D$ is the dominance parameter defined in
Eq.~(\ref{eq:dominance-parameter}). Consequently,
$D=1$ corresponds to
$\operatorname{Br}(A'\to a\gamma)=1/2$.

\begin{figure}[t]
\centering
\includegraphics[width=0.98\linewidth]
{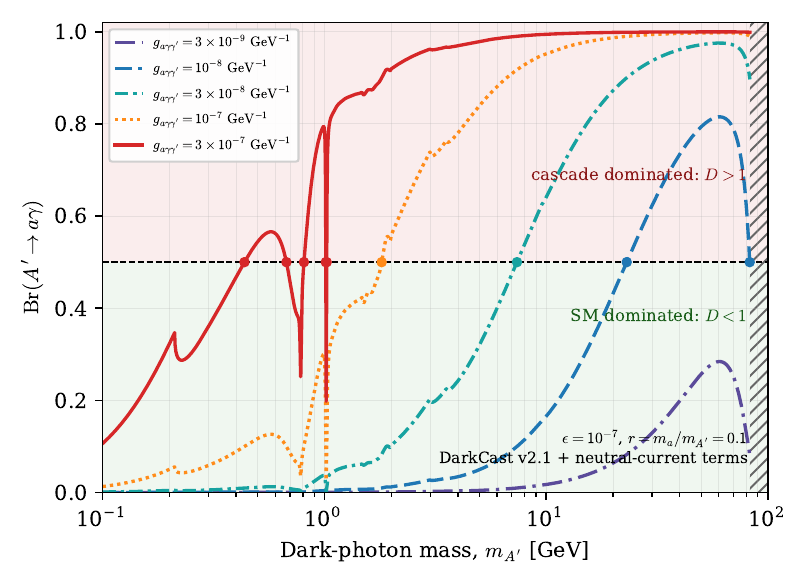}
\caption{Dark-photon branching fraction
$\operatorname{Br}(A'\to a\gamma)$ as a function of $m_{A'}$ for
$\epsilon=10^{-7}$ and $r=m_a/m_{A'}=0.1$. The Standard Model width is
the DarkCast v2.1 width (mass-dependent leptonic and hadronic
contributions) plus the neutral-current and $A'\to\nu\bar\nu$ terms of
Sec.~\ref{subsec:dark-photon-widths}. The horizontal dashed line corresponds to
$D=1$, or equivalently
$\operatorname{Br}(A'\to a\gamma)=1/2$, and the markers indicate the
crossing points. The upper and lower shaded regions correspond to
cascade-dominated and Standard Model-dominated decays, respectively.
The hatched region $m_{A'}\geq82.9~{\rm GeV}$ is not an experimental
exclusion; it is kinematically inaccessible through on-shell
$Z\to aA'$ production for $r=0.1$.}
\label{fig:branching-ratio-fcc}
\end{figure}

As shown in Fig.~\ref{fig:branching-ratio-fcc}, increasing
$g_{a\gamma\gamma'}$ shifts the transition to the cascade-dominated
regime toward lower dark-photon masses. This behavior follows from the
different mass dependence of the two contributions. For a fixed mass
ratio $r$, the exotic width scales approximately as
$\Gamma(A'\to a\gamma)\propto
g_{a\gamma\gamma'}^{2}m_{A'}^{3}$, whereas the Standard Model width
scales approximately as
$\Gamma_{\rm SM}(A')\propto\epsilon^{2}m_{A'}$, apart from threshold and
hadronic-resonance effects and the neutral-current growth close to $m_Z$. Therefore, the dominance parameter grows
approximately as
$D\propto g_{a\gamma\gamma'}^{2}m_{A'}^{2}/\epsilon^{2}$.

The $D=1$ transitions of the smaller couplings are
\begin{equation}
\begin{aligned}
g_{a\gamma\gamma'}=10^{-8}~{\rm GeV}^{-1}
&:\quad m_{A'}\simeq23.07~{\rm GeV}\ \text{and}\\
&\phantom{:}\quad m_{A'}\simeq82.71~{\rm GeV},\\
g_{a\gamma\gamma'}=3\times10^{-8}~{\rm GeV}^{-1}
&:\quad m_{A'}\simeq7.38~{\rm GeV},\\
g_{a\gamma\gamma'}=10^{-7}~{\rm GeV}^{-1}
&:\quad m_{A'}\simeq1.81~{\rm GeV}.
\end{aligned}
\label{eq:dominance-transition-masses}
\end{equation}
For $g_{a\gamma\gamma'}=10^{-8}~{\rm GeV}^{-1}$ the second crossing at
$82.71~{\rm GeV}$ is a downward one: close to the kinematic endpoint the
neutral-current terms make $\Gamma_{\rm SM}$ grow faster than
$m_{A'}^3$, and the Standard Model width dominates again. For
$g_{a\gamma\gamma'}=3\times10^{-9}~{\rm GeV}^{-1}$ there is no crossing
at all; $D$ reaches its maximum, $0.40$, at $m_{A'}\simeq60.5~{\rm GeV}$.
With the electromagnetic width alone this curve would cross $D=1$ at
$75.84~{\rm GeV}$, and the other three transitions would lie at
$22.75$, $7.38$, and $1.81~{\rm GeV}$.
For $g_{a\gamma\gamma'}=3\times10^{-7}~{\rm GeV}^{-1}$, the
mass-dependent hadronic width produces several crossings near
$m_{A'}\simeq0.44$--$1.02~{\rm GeV}$. These multiple crossings reflect
the physical hadronic-threshold and vector-meson resonance structure
contained in $\Gamma_{\rm SM}(A')$; they are not numerical artifacts.
Away from this resonance region, the exotic decay rapidly becomes
dominant.

The branching fraction alone does not determine whether the
five-photon signature is observable. The decay positions of $A'$,
$a_1$, and $a_2$, as well as the detector acceptance and photon
reconstruction efficiencies, must also be included. These effects are
considered in the following subsections.

\subsection{Dark-Photon Decay Length and Fiducial Probability}
\label{subsec:dark-photon-fiducial-probability}

The branching fraction discussed in the previous subsection determines
which dark-photon decay channel is dominant, but it does not determine
whether the decay takes place inside the detector. The spatial
distribution of the dark-photon decays is controlled by its mean
laboratory decay length,
\begin{equation}
L_{A'}
=
\beta_{A'}\gamma_{A'}
\frac{\hbar c}{\Gamma_{A'}^{\mathrm{tot}}},
\label{eq:numerical-dark-photon-decay-length}
\end{equation}
where $\Gamma_{A'}^{\mathrm{tot}}$ is given in
Eq.~(\ref{eq:dark-photon-total-width}). The boost factor
$\beta_{A'}\gamma_{A'}$ is calculated from the two-body kinematics of
$Z\to aA'$ using Eq.~(\ref{eq:z-decay-final-state-boosts}). Therefore,
no constant or externally assigned boost factor is used in the present
analysis.

For a particle with mean laboratory decay length $L_{A'}$, the
probability of decaying between the path lengths $\ell_{\min}$ and
$\ell_{\max}$ is
\begin{equation}
P_{\mathrm{fid}}^{A'}
\left(\ell_{\min},\ell_{\max}\right)
=
\exp\left(-\frac{\ell_{\min}}{L_{A'}}\right)
-
\exp\left(-\frac{\ell_{\max}}{L_{A'}}\right),
\label{eq:dark-photon-fiducial-probability}
\end{equation}
as introduced generally in
Eq.~(\ref{eq:fiducial-decay-probability}). We adopt the reference
fiducial interval
\begin{equation}
\ell_{\min}=1~\mathrm{mm},
\qquad
\ell_{\max}=2~\mathrm{m}.
\label{eq:reference-fiducial-window}
\end{equation}
The lower boundary represents an illustrative minimum displacement
required to separate a displaced decay from the interaction point,
whereas the upper boundary is motivated by the approximate radial
extent of the IDEA central tracking system proposed for the
FCC-ee~\cite{IDEADetector}. This interval should be regarded as a
detector-scale reference window rather than a complete detector
simulation. In particular, the actual path-length boundaries depend on
the polar angle of the particle and on the detailed detector geometry.

\begin{figure*}[t]
\centering
\includegraphics[width=0.96\textwidth]
{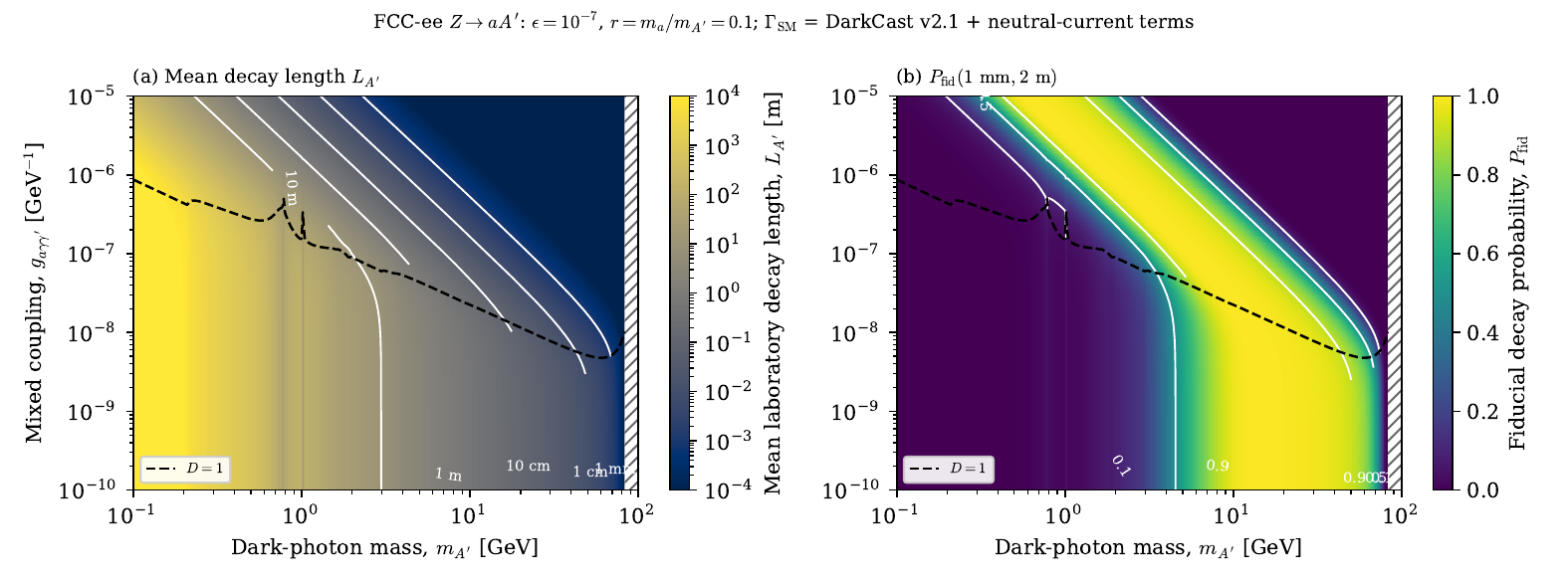}
\caption{Dark-photon decay length and fiducial decay probability in the
$(m_{A'},g_{a\gamma\gamma'})$ plane for $\epsilon=10^{-7}$ and
$r=m_a/m_{A'}=0.1$. Panel~(a) shows the mean laboratory decay length
$L_{A'}$, calculated using the mass-dependent boost from
$Z\to aA'$ and the total dark-photon width, whose Standard Model part
is the DarkCast v2.1 width plus the neutral-current and
$A'\to\nu\bar\nu$ terms of Sec.~\ref{subsec:dark-photon-widths}. The white contours correspond to
$L_{A'}=1~\mathrm{mm}$, $1~\mathrm{cm}$, $10~\mathrm{cm}$,
$1~\mathrm{m}$, and $10~\mathrm{m}$. Panel~(b) shows the probability
$P_{\mathrm{fid}}^{A'}$ for the dark photon to decay within the
reference interval $1~\mathrm{mm}<\ell<2~\mathrm{m}$. The labeled
contours correspond to $P_{\mathrm{fid}}^{A'}=0.1$, $0.5$, and $0.9$.
The black dashed curve denotes the $D=1$ boundary, at which the exotic
and Standard Model dark-photon widths are equal. The hatched region
$m_{A'}\geq82.9~\mathrm{GeV}$ is kinematically inaccessible through
on-shell $Z\to aA'$ production for $r=0.1$.}
\label{fig:dark-photon-decay-fcc}
\end{figure*}

Panel~(a) of Fig.~\ref{fig:dark-photon-decay-fcc} shows that the
dark-photon decay length decreases as either $m_{A'}$ or
$g_{a\gamma\gamma'}$ is increased. This behavior results from the
growth of the total width with both the dark-photon mass and the
mixed-portal coupling. In addition, the boost factor decreases as the
kinematic boundary $m_a+m_{A'}=m_Z$ is approached. Consequently,
dark photons produced close to the upper mass boundary have smaller
laboratory decay lengths than would be obtained using a fixed boost
approximation. The neutral-current terms shorten $L_{A'}$ further where
$\Gamma_{\rm SM}$ dominates. For $g_{a\gamma\gamma'}=10^{-9}~\mathrm{GeV}^{-1}$
the decay length is reduced by factors $0.95$, $0.65$, and $0.14$ at
$m_{A'}=30$, $60$, and $80~\mathrm{GeV}$, relative to the
electromagnetic width alone. At the benchmark the change is
$4\times10^{-4}$.

The color scale in panel~(b) represents the fraction of produced dark
photons that decay between $1~\mathrm{mm}$ and $2~\mathrm{m}$. Thus,
the contours labeled $0.1$, $0.5$, and $0.9$ correspond to
fiducial decay probabilities of $10\%$, $50\%$, and $90\%$,
respectively. These values are decay-position probabilities and should
not be confused with the branching fraction
$\operatorname{Br}(A'\to a\gamma)$ shown in
Fig.~\ref{fig:branching-ratio-fcc}.

Each probability value appears on two different contour branches. On
the low-mass or long-lifetime side of the high-probability band, the
dark photon typically travels beyond $\ell_{\max}=2~\mathrm{m}$ before
decaying. On the high-mass or short-lifetime side, it typically decays
before reaching the minimum displacement
$\ell_{\min}=1~\mathrm{mm}$. The central yellow band identifies the intermediate-lifetime region in which most dark photons
decay inside the selected fiducial volume.

For the reference interval of
Eq.~(\ref{eq:reference-fiducial-window}), the fiducial probability is
maximized when
\begin{equation}
L_{A'}^{\mathrm{opt}}
=
\frac{\ell_{\max}-\ell_{\min}}
{\ln\left(\ell_{\max}/\ell_{\min}\right)}
\simeq 0.263~\mathrm{m}.
\label{eq:optimal-dark-photon-decay-length}
\end{equation}
At this value, the maximum probability is approximately
\begin{equation}
P_{\mathrm{fid,max}}^{A'}
\simeq 0.996.
\label{eq:maximum-dark-photon-fiducial-probability}
\end{equation}
This explains why the central part of the yellow band approaches unity,
while the probability decreases on both sides of the band.

The black dashed $D=1$ curve provides complementary information.
It separates the Standard Model-dominated and cascade-dominated decay
regimes, whereas the colored regions determine whether the dark photon
decays inside the selected detector volume. A parameter point favorable
for the observable five-photon signal must therefore have both a
sufficiently large exotic branching fraction and a sufficiently large
fiducial decay probability. At the particle level, the corresponding
dark-photon contribution to the observable cascade rate is proportional
to
\begin{equation}
\operatorname{Br}(A'\to a\gamma)
P_{\mathrm{fid}}^{A'}.
\label{eq:dark-photon-observable-cascade-factor}
\end{equation}
The decay probabilities of the two ALPs, together with the geometrical
acceptance and photon-reconstruction efficiencies, must be included
separately in the complete five-photon event yield.

\subsection{ALP Decay Lengths in the Cascade}
\label{subsec:alp-decay-lengths}

Although $a_1$ and $a_2$ represent the same ALP species and therefore have the same proper decay width, their laboratory-frame decay lengths are generally different because they originate from different stages of the cascade. The directly produced ALP, $a_1$, acquires its boost from the two-body decay $Z\to a_1A'$, whereas the secondary ALP, $a_2$, is produced through the subsequent decay $A'\to a_2\gamma$.

The ALP diphoton width and total width are given in Eqs.~(\ref{eq:alp-diphoton-width}) and (\ref{eq:alp-total-width}), respectively. The laboratory decay length of $a_1$ is consequently determined by Eq.~(\ref{eq:dark-photon-direct-alp-decay-lengths}). For the secondary ALP, both the boost and the decay length depend on the decay angle $\theta^{*}$ in the dark-photon rest frame, as shown in Eq.~(\ref{eq:secondary-alp-boost-and-decay-length}).

\begin{figure*}[t]
\centering
\includegraphics[width=0.96\textwidth]{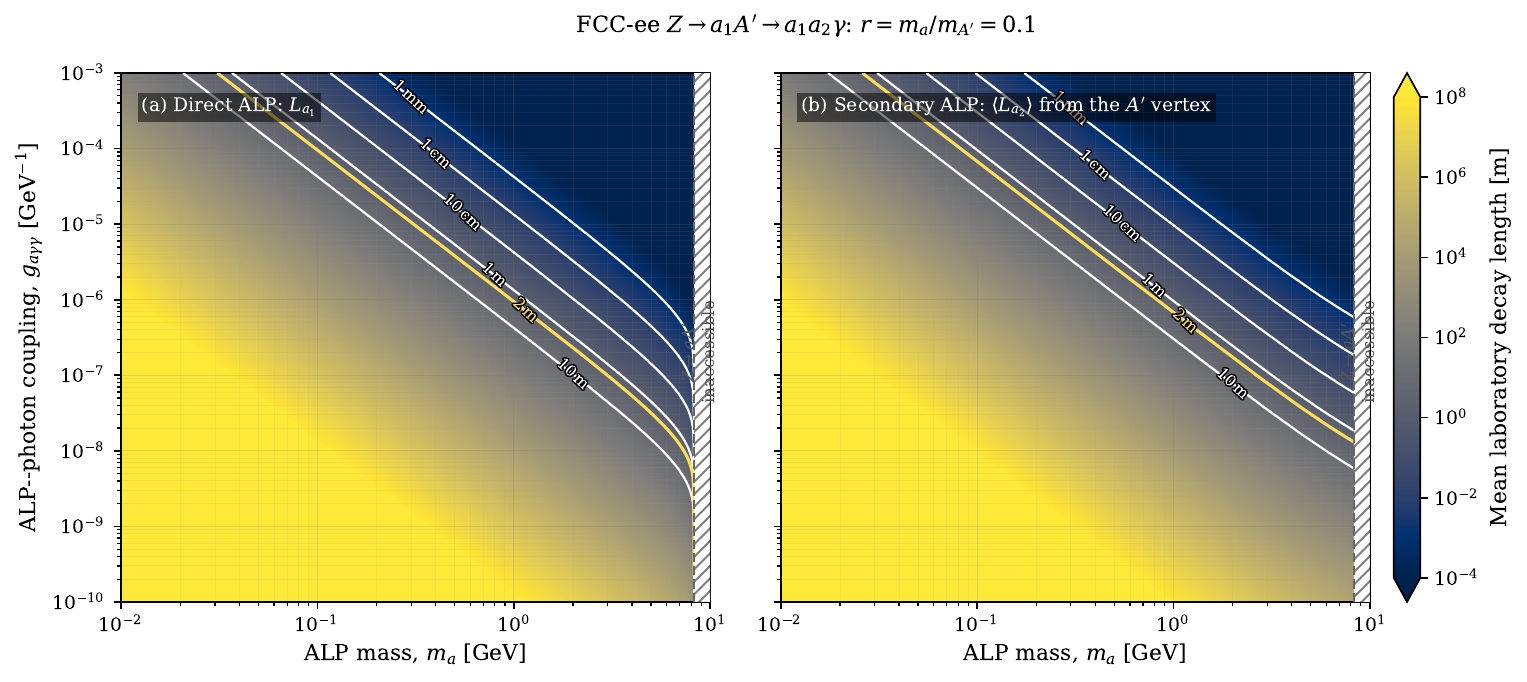}
\caption{
Mean laboratory decay lengths of the ALPs in the cascade
$Z\to a_1A'\to a_1a_2\gamma$ for the reference mass ratio
$r=m_a/m_{A'}=0.1$.
The left panel shows the decay length $L_{a_1}$ of the ALP produced directly in the $Z$-boson decay.
The right panel shows the angular-averaged decay length
$\langle L_{a_2}\rangle$ of the secondary ALP, measured from the $A'$ decay vertex.
The white contours correspond to mean decay lengths of
$1,\mathrm{mm}$, $1,\mathrm{cm}$, $10,\mathrm{cm}$,
$1,\mathrm{m}$, and $10,\mathrm{m}$.
The yellow contour indicates the reference outer-tracker scale of $2,\mathrm{m}$.
The hatched region is kinematically inaccessible to the on-shell decay
$Z\to aA'$.
The displayed parameter plane illustrates the lifetime regimes and should not be interpreted as an experimentally allowed region; the existing constraints on $g_{a\gamma\gamma}$ must be applied using Fig.~\ref{fig:alp-photon-constraints}.
}
\label{fig:alp-decay-fcc}
\end{figure*}

For the lifetime map shown in Fig.~\ref{fig:alp-decay-fcc}, we assume an
isotropic decay in the $A'$ rest frame and define the angular-averaged mean
flight length of $a_2$ as
\begin{equation}
\left\langle L_{a_2}\right\rangle
=
\frac{1}{2}
\int_{-1}^{1}
L_{a_2}(\theta^{*})\,
\mathrm{d}\!\left(\cos\theta^{*}\right).
\label{eq:secondary-alp-angular-averaged-length}
\end{equation}
This quantity is measured from the dark-photon decay vertex rather than from the interaction point. The angular average provides a transparent kinematic estimate, while a complete detector simulation would additionally retain the spin correlations and detailed detector geometry.

For a fixed ALP mass, both decay lengths scale approximately as
$g_{a\gamma\gamma}^{-2}$. Therefore, sufficiently large values of
$g_{a\gamma\gamma}$ lead to prompt diphoton decays, whereas smaller couplings produce displaced decays or ALPs that escape the reference detector volume.

The direct and secondary ALPs exhibit different lifetime contours because their boosts are determined by different stages of the cascade. As the boundary $m_a+m_{A'}=m_Z$ is approached, the momentum of the directly produced $a_1$ tends to zero. By contrast, $a_2$ retains a nonzero momentum originating from the decay $A'\to a_2\gamma$, even when the parent dark photon is produced nearly at rest.

The right panel of Fig.~\ref{fig:alp-decay-fcc} gives the flight length of $a_2$ relative to its production vertex and does not by itself determine the displacement relative to the interaction point. The complete five-photon acceptance must also include the preceding dark-photon flight distance. Accordingly, the observable event yield requires a sequential treatment of the $A'$ decay probability and the subsequent decay probabilities of both ALPs.

\subsection{Sequential Five-Photon Yield at FCC-ee}
\label{subsec:five-photon-yield}

We estimate the expected five-photon yield from the sequential decay chain
\begin{equation}
Z \to a_{1}A' \to a_{1}a_{2}\gamma \to 5\gamma .
\label{eq:five-photon-chain-yield}
\end{equation}
The number of events for which the required decays occur inside the fiducial detector volume is
\begin{equation}
\begin{aligned}
N_{5\gamma}^{\mathrm{fid}}
={}&
N_Z\,
\mathrm{Br}(Z\to aA')\,
\mathrm{Br}(A'\to a\gamma)
\\
&\times
\left[
\mathrm{Br}(a\to\gamma\gamma)
\right]^2
P_{\mathrm{seq}}(R),
\end{aligned}
\label{eq:five-photon-fiducial-yield}
\end{equation}

Here, $N_Z$ is the number of $Z$ bosons produced at the interaction point associated with the detector under consideration, and $P_{\mathrm{seq}}(R)$ is the sequential decay probability for a detector with fiducial radius $R$. Although the complete FCC-ee program is expected to produce approximately $6\times10^{12}$ $Z$ bosons across four interaction points, the fiducial estimate of this subsection is given for a single detector. We therefore use

\begin{equation}
N_Z=1.5\times10^{12}
\end{equation}
and adopt a spherical reference volume with $R=2~\mathrm{m}$. The yields of this subsection should consequently be interpreted as single-detector estimates; the detector-level analysis of Sec.~\ref{subsec:detector-level-analysis} quotes yields for one and for four interaction points.

The primary ALP $a_1$ is produced at the interaction point, whereas the secondary ALP $a_2$ is produced at the dark-photon decay vertex. The sequential decay probability is therefore evaluated as

\begin{equation}
P_{\mathrm{seq}}(R)
=
P_{a_1}(R)\,
I_{A'a_2}(R).
\label{eq:sequential-decay-probability}
\end{equation}

\begin{equation}
P_{a_1}(R)
=
1-\exp\left(-\frac{R}{L_{a_1}}\right).
\label{eq:primary-alp-decay-probability}
\end{equation}

\begin{equation}
I_{A'a_2}(R)
=
\int_{0}^{R}
\frac{\mathrm{d}s}{L_{A'}}
\exp\left(-\frac{s}{L_{A'}}\right)
\overline{P}_{a_2}(s).
\label{eq:dark-photon-secondary-alp-probability}
\end{equation}

\begin{equation}
\overline{P}_{a_2}(s)
=
\frac{1}{2}
\int_{-1}^{1}
\mathrm{d}(\cos\theta^{*})
\left[
1-
\exp\left(
-\frac{d_{\max}(s,\theta_{\mathrm{lab}})}
{L_{a_2}(\theta^{*})}
\right)
\right].
\label{eq:secondary-alp-angular-probability}
\end{equation}

Here, $P_{a_1}(R)$ is the probability that the primary ALP decays before leaving the fiducial volume. The quantity $I_{A'a_2}(R)$ accounts for the dark-photon decay position and the subsequent decay of the secondary ALP. The variable $s$ is the dark-photon decay distance, while $\theta^{*}$ is the emission angle of $a_2$ in the dark-photon rest frame. The remaining path length of $a_2$ inside the spherical fiducial volume is
\begin{equation}
d_{\max}(s,\theta_{\mathrm{lab}})
=
-s\cos\theta_{\mathrm{lab}}
+
\sqrt{
R^{2}
-
s^{2}\sin^{2}\theta_{\mathrm{lab}}
}.
\label{eq:secondary-alp-remaining-distance}
\end{equation}
The laboratory-frame angle $\theta_{\mathrm{lab}}$ and the mean decay length $L_{a_2}(\theta^{*})$ are obtained from the exact two-body kinematics of $Z\to a_1A'$ and $A'\to a_2\gamma$.

We consider the representative benchmark point
\begin{equation}
m_{A'}=10~\mathrm{GeV},
\qquad
m_a=1~\mathrm{GeV},
\qquad
\epsilon=10^{-7},
\label{eq:five-photon-benchmark-masses}
\end{equation}
with
\begin{equation}
g_{a\gamma\gamma'}=10^{-6}~\mathrm{GeV}^{-1},
\qquad
g_{a\gamma\gamma}=10^{-6}~\mathrm{GeV}^{-1}.
\label{eq:five-photon-benchmark-couplings}
\end{equation}
This point is not excluded by the direct ALP and LEP constraints displayed in the parameter-space figures used in this work. This observation should not be interpreted as a complete global recast of all beam-dump, displaced, and missing-energy searches.

The numerical results are summarized in Table~\ref{tab:five-photon-benchmark}.

\begin{table}[t]
\centering
\caption{Numerical results for the representative five-photon benchmark. The event yield includes the sequential decay probability inside $R=2~\mathrm{m}$, but does not include photon-reconstruction efficiencies or analysis cuts.}
\label{tab:five-photon-benchmark}
\begin{tabular}{ll}
$\Gamma(Z\to aA')$
&
$7.29\times10^{-10}~\mathrm{GeV}$
\\
$\mathrm{Br}(Z\to aA')$
&
$2.92\times10^{-10}$
\\
$N_Z\times \mathrm{Br}(Z\to aA')$
&
$4.38\times10^{2}$
\\
$\Gamma(A'\to a\gamma)$
&
$3.22\times10^{-12}~\mathrm{GeV}$
\\
$\mathrm{Br}(A'\to a\gamma)$
&
$0.9995$
\\
$\Gamma(a\to\gamma\gamma)$
&
$4.97\times10^{-15}~\mathrm{GeV}$
\\
$L_{A'}$
&
$2.76\times10^{-4}~\mathrm{m}$
\\
$L_{a_1}$
&
$1.79~\mathrm{m}$
\\
$P_{\mathrm{seq}}(R=2~\mathrm{m})$
&
$0.586$
\\
$N_{5\gamma}^{\mathrm{fid}}$
&
$2.57\times10^{2}$
\end{tabular}
\end{table}

Before the fiducial decay requirement, the expected number of cascade events is

\begin{equation}
\begin{aligned}
N_{5\gamma}^{\mathrm{cascade}}
={}&
N_Z\,
\mathrm{Br}(Z\to aA')\,
\mathrm{Br}(A'\to a\gamma)
\\
&\times
\left[
\mathrm{Br}(a\to\gamma\gamma)
\right]^2 .
\end{aligned}
\label{eq:five-photon-cascade-yield}
\end{equation}

\begin{equation}
N_{5\gamma}^{\mathrm{cascade}}
\simeq
438 .
\label{eq:five-photon-cascade-yield-numerical}
\end{equation}

After including the sequential decay probability, we obtain
\begin{equation}
N_{5\gamma}^{\mathrm{fid}}
\simeq
438
\times
0.586
\simeq
257.
\label{eq:five-photon-numerical-yield}
\end{equation}

For this benchmark, the primary-ALP decay probability inside $R=2~\mathrm{m}$ is approximately $0.673$, while the conditional probability that the dark photon and the secondary ALP complete their decays inside the remaining fiducial volume is approximately $0.870$. The dark photon is predominantly prompt, with $L_{A'}\simeq0.276~\mathrm{mm}$. A more restrictive requirement that all three decay vertices occur beyond $1~\mathrm{mm}$ gives approximately $7$ events. We therefore use the inclusive fiducial definition in Eq.~(\ref{eq:five-photon-fiducial-yield}).

The approximate minimum diphoton opening angles are $44.4~\mathrm{mrad}$ for $a_1$ and $43.9~\mathrm{mrad}$ for $a_2$. These estimates indicate that the two photons from an ALP decay are not necessarily fully merged at this benchmark point. Nevertheless, detector granularity, photon clustering, energy thresholds, and angular acceptance must be included in a dedicated detector-level analysis.

The reconstructed event yield can be parametrized as
\begin{equation}
N_{5\gamma}^{\mathrm{obs}}
=
N_{5\gamma}^{\mathrm{fid}}
\mathcal{A}_{5\gamma}
\epsilon_{\gamma}^{5}.
\label{eq:five-photon-observed-yield}
\end{equation}
where $\mathcal{A}_{5\gamma}$ is the total kinematic and geometric acceptance and $\epsilon_{\gamma}$ is the single-photon reconstruction efficiency. The quantity $N_{5\gamma}^{\mathrm{fid}}\simeq257$ should therefore be regarded as a fiducial preselection estimate. In the following subsection, we replace this schematic acceptance factor by an event-by-event detector-oriented treatment of the five-photon final state.

\subsection{Detector-Level Five-Photon Analysis}
\label{subsec:detector-level-analysis}

\subsubsection{IDEA Geometry and Event Reconstruction}
\label{subsubsec:idea-geometry}

To connect the fiducial estimate above with an experimentally
reconstructable signature, we perform a fast detector-oriented analysis
of the complete cascade
\begin{equation}
e^{+}e^{-}\to Z\to a_{1}A'
\to a_{1}a_{2}\gamma
\to 5\gamma
\end{equation}
at the FCC-ee $Z$ pole. Signal events are generated with
MadGraph5\_aMC@NLO v3.7.2~\cite{MadGraph5aMC} using the
\texttt{DarkALPPortal\_UFO} model implemented in the UFO
format~\cite{UFO}.
For the benchmark point of
Eqs.~(\ref{eq:five-photon-benchmark-masses}) and
(\ref{eq:five-photon-benchmark-couplings}), the event sample is
generated at $\sqrt{s}=91.1876~\mathrm{GeV}$, corresponding to beam
energies of $45.5938~\mathrm{GeV}$. The validated sample contains
$10^4$ events, with exactly five final-state photons in each event. No
beam-energy convolution or QED ISR photon is included in this nominal
sample.

The generated four-vectors provide an independent check of the
two-body kinematics derived above. For the benchmark masses, the
reconstructed mean boost factors are
\begin{equation}
\langle\beta\gamma\rangle_{A'}=4.50,
\qquad
\langle\beta\gamma\rangle_{a_1}=45.04,
\qquad
\langle\beta\gamma\rangle_{a_2}=23.29.
\end{equation}
Using the benchmark widths, these correspond to mean laboratory decay
lengths of approximately
\begin{equation}
\begin{aligned}
\langle L_{A'}\rangle&=2.76\times10^{-4}~\mathrm{m},
\\
\langle L_{a_1}\rangle&=1.79~\mathrm{m},
\\
\langle L_{a_2}\rangle&=0.92~\mathrm{m}.
\end{aligned}
\end{equation}
The first two values agree with the analytic benchmark estimates above,
while the secondary-ALP value reflects its event-by-event boost
distribution in the cascade.

\paragraph{ISR and beam-energy-spread stress test.}
The published FCC-ee reference parameters at the $Z$ pole give a
natural single-beam relative RMS energy spread of $0.038\%$
($17~\mathrm{MeV}$) and a collision value of $0.132\%$ after
beam--beam effects~\cite{FCCEnergyCalibration}.  We use the latter as a
conservative Gaussianized single-beam width and draw the two beam
energies independently.  This gives an RMS spread of approximately
$84.8~\mathrm{MeV}$ in the hard-process center-of-mass energy in the
beam-spread-only test.

For ISR, we use the leading soft-collinear structure-function radiator
\cite{FrixioneISR2022}
\begin{equation}
\begin{aligned}
H(z)&=\beta_{\rm ISR}(1-z)^{\beta_{\rm ISR}-1},
\\
\beta_{\rm ISR}
&=
\frac{2\alpha}{\pi}
\left[\ln\!\left(\frac{s}{m_e^2}\right)-1\right]
\\
&=0.1077,
\end{aligned}
\label{eq:isr-leading-radiator}
\end{equation}
where $z=s'/s$ is the fraction of the nominal squared collision energy
remaining in the hard process.  Each draw is weighted by the $Z$
Breit--Wigner line shape and the relative $p_*^3$ two-body phase-space
factor.  A one-sided collinear recoil, with a randomly selected emitting
beam, is used to obtain a conservative longitudinal boost of the hard
system.  For every morphed event, the original decay angles are retained,
the complete two-body cascade is rebuilt at $\sqrt{s'}$, and the
sequential decay positions and the full IDEA truth-level selection are
re-evaluated.  The test uses the $10^4$ parent events, 20 decay replicas
per event, and 24 independent ISR/beam-spectrum draws.  The fixed-energy
reconstruction gives $\mathcal{A}_{\rm sel}=0.3071$, consistent with the
quoted nominal value.

\begin{table*}[t]
\centering
\caption{Truth-level kinematic stress test for the benchmark
$(m_{A'},m_a)=(10,1)~\mathrm{GeV}$.  The ISR rows are normalized to
produced resonant events through the Breit--Wigner and phase-space
weights.  The final column is the relative change with respect to the
fixed-energy acceptance in this test.}
\label{tab:isr-beam-stress-test}
\scriptsize
\setlength{\tabcolsep}{3pt}
\begin{tabular}{lccccc}
Scenario
& $\langle\sqrt{s'}\rangle\ (\sigma)$
& $\langle|\beta_z|\rangle$
& $\langle\beta\gamma\rangle_{A'}$
& $\langle\beta\gamma\rangle_{a_1}$
& $\mathcal{A}_{\rm sel}$ [$\Delta\mathcal{A}/\mathcal{A}$]
\\
& [GeV] & & & &
\\ \hline
Fixed energy
& 91.1876 (0) & 0 & 4.504 & 45.04 & 0.3071 [reference]
\\
Beam spread
& 91.188 (0.0848) & $7.45\times10^{-4}$
& 4.504 & 45.04 & 0.3071 [$-0.014\%$]
\\
Leading-log ISR
& 91.000 (0.681) & $2.09\times10^{-3}$
& 4.495 & 44.95 & 0.3073 [$+0.050\%$]
\\
ISR plus beam spread
& 91.000 (0.687) & $2.61\times10^{-3}$
& 4.495 & 44.95 & 0.3072 [$+0.042\%$]
\end{tabular}
\end{table*}

The beam spread alone leaves the mean direct-particle boosts unchanged
to better than $10^{-3}$ and produces relative RMS widths of about
$0.11\%$ in $(\beta\gamma)_{A'}$ and
$(\beta\gamma)_{a_1}$.  With ISR and the beam spread combined, the RMS
widths become $0.040$ and $0.392$ in the two boost factors,
respectively, while their means decrease by about $0.21\%$.  The
secondary-ALP boost already has a broad decay-angle distribution; its
mean changes from $23.29$ to $23.26$ ($-0.17\%$), with essentially
no change to its intrinsic width.  The corresponding mean decay lengths
are
\begin{equation}
\begin{aligned}
\langle L_{A'}\rangle&=2.756\times10^{-4}~\mathrm{m},
\\
\langle L_{a_1}\rangle&=1.783~\mathrm{m},
\\
\langle L_{a_2}\rangle&=0.923~\mathrm{m},
\end{aligned}
\end{equation}
compared with $2.761\times10^{-4}~\mathrm{m}$, $1.787~\mathrm{m}$,
and $0.924~\mathrm{m}$ in the nominal sample.  The central combined
acceptance shift is only $+0.04\%$; the trial-to-trial RMS of the
relative shift is $0.79\%$, so no acceptance change is resolved at the
percent level.  We therefore retain the nominal benchmark yield and treat $1\%$ as the numerical resolution of this particular stress test, not as a calibrated systematic uncertainty. This test uses the non-pointing selection of Sec.~\ref{subsubsec:nonpointing} with $E_\gamma>2~\mathrm{GeV}$, $d_{\min}^{\rm ECAL}>20~\mathrm{mm}$, and $\theta_{\rm NP}^{\max}>8~\mathrm{mrad}$, whose nominal acceptance is $0.307$. For the DV selection of Sec.~\ref{subsubsec:dv-selection}, the same ISR and beam-spread morphing changes the signal efficiency of each requirement by less than $0.5\%$.

This procedure does not generate resolved wide-angle ISR photons, model
ISR--final-state-radiation interference, or replace the Gaussianized
beam spectrum by a machine simulation with correlations and non-Gaussian
tails.  A dedicated ISR-enabled event generation convolved with the
final FCC-ee luminosity spectrum remains necessary for an experimental
projection.  Since $N_Z$ in Eq.~(\ref{eq:number-of-produced-a-dark-photon-pairs})
denotes the number of actually produced resonant $Z$ bosons, no
additional ISR suppression is applied to the production normalization;
such a correction would instead be required if the rate were normalized
directly from integrated luminosity and a Born cross section.

For the detector-geometry study, each generated event is reused with
20 independent stochastic decay realizations, giving
$2.0\times10^{5}$ decay trials. The sequential positions of the
$A'$, $a_1$, and $a_2$ decays are generated using their
event-dependent laboratory boosts and decay lengths, so that the
secondary-ALP trajectory begins at the simulated dark-photon decay
vertex rather than at the interaction point.

The IDEA electromagnetic calorimeter is represented by a cylindrical
active volume with barrel radius
\begin{equation}
R_{\rm ECAL}=2.25~\mathrm{m},
\end{equation}
half-length
\begin{equation}
Z_{\rm ECAL}=2.40~\mathrm{m},
\end{equation}
and a circular endcap beam hole of radius
\begin{equation}
R_{\rm hole}=0.36~\mathrm{m}.
\end{equation}
These dimensions follow the current parameterized IDEA crystal-ECAL
geometry, which has a nominal crystal front-face size of
$10\times10~\mathrm{mm}^{2}$ and uses PbWO$_4$ as the baseline crystal
material, with a Moli\`ere radius of approximately $20~\mathrm{mm}$
\cite{IDEADetector}.
Photons are propagated in straight lines from their actual production
vertices to the first intersection with the active ECAL barrel or
endcap surface. Their ECAL impact positions are then used to determine
geometrical acceptance, pairwise cluster separations, and non-pointing
angles. This procedure constitutes a fast truth-level detector model
rather than a full Geant-based simulation; the detector effects most
relevant for the present five-photon topology are introduced explicitly
through the ECAL geometry, photon-energy thresholds, cluster-separation
requirements, and angular-resolution model discussed below. In
particular, the calorimeter surfaces are treated as continuous: crystal
boundaries, electromagnetic-shower development, digitization, and a
two-shower clustering algorithm are not simulated.

For comparison with the spherical reference volume used in
Sec.~\ref{subsec:five-photon-yield}, the same decay simulation gives a
complete-chain probability of about $0.58$ inside a sphere of radius
$2~\mathrm{m}$. In the cylindrical IDEA geometry, the corresponding
probability that the complete cascade occurs before the relevant
particles leave the detector volume increases to approximately
\begin{equation}
P_{\rm chain}^{\rm IDEA}=0.703.
\end{equation}
The remaining loss of signal events is controlled not only by
the decay geometry but also by ECAL acceptance, photon merging, energy
requirements, and the displaced-photon selections considered in the
following subsections.

\paragraph{Photon response and event samples.}
Each photon is assigned a measured energy and a measured impact
position. The crystal section of a dual-readout calorimeter of this
type has been simulated in Ref.~\cite{Lucchini2022}, which reports
\begin{equation}
\frac{\sigma_E}{E}=\frac{2.5\%}{\sqrt{E}}\oplus0.6\%,
\qquad
\sigma_\theta=\frac{1.56}{\sqrt{E}}\oplus0.33~\mathrm{mrad},
\label{eq:lucchini-resolution}
\end{equation}
with $E$ in GeV. The second expression is the resolution of the shower
position, i.e., of the polar angle of the cluster seen from the
interaction point, not of the shower direction. No $\phi$ resolution
is given; we assume that the angular resolution is the same in the two
directions transverse to the line of sight, $\sigma_\phi\sin\theta=
\sigma_\theta$. The angular resolutions are applied unchanged at our
calorimeter radius. For photons with a flat $5$--$100~\mathrm{GeV}$
spectrum, Eq.~(\ref{eq:lucchini-resolution}) gives an RMS of
$0.43~\mathrm{mrad}$, close to the $0.41~\mathrm{mrad}$ found in the full
simulation of the present IDEA crystal geometry~\cite{IDEADetector}.
Varying the $\phi$ assumption by factors of 2, using a fixed linear
instead of a fixed angular resolution, or replacing the energy term by
the IDEA parametrization $3.0\%/\sqrt{E}\oplus0.5\%$~\cite{IDEADetector}
changes the selected signal yield of Sec.~\ref{subsubsec:dv-selection}
by less than $0.5\%$.

The yields refer to the FCC-ee $Z$-pole program of
$205~\mathrm{ab}^{-1}$~\cite{FCC2025,Polesello2025}, split into
$40$, $125$, and $40~\mathrm{ab}^{-1}$ at $\sqrt{s}=87.9$, $91.19$, and $94.3~\mathrm{GeV}$,
and equally shared by four interaction points (IPs). The signal is normalized
at each energy with the ISR-convolved resonance line shape; the
off-peak points add $15.4$ and $35.2$ produced cascades per interaction point to the
$390.8$ produced on the peak. The detector-level acceptance at the
off-peak energies is obtained by rebuilding the benchmark events at the
corresponding $\sqrt{s}$, and the combined yields differ from a pure
on-peak scaling by the factor $\rho\simeq1.007$. The prompt
$e^+e^-\to5\gamma$ background is generated with MadGraph5\_aMC@NLO at
leading order at the three energies, with photons satisfying
$E_\gamma>1~\mathrm{GeV}$, $|\eta|<3$, and $\Delta R>0.01$. These
generator cuts are looser than the detector acceptance, whose endcap
hole corresponds to $|\eta|\simeq2.60$.

\subsubsection{ECAL Photon Separation and Five-Cluster Resolution}
\label{subsubsec:ecal-separation}

A characteristic feature of the signal is that the two photons from an
ALP decay can be moderately collimated in the laboratory frame.  To
quantify whether the complete five-photon final state can be resolved in
the electromagnetic calorimeter, we define
\begin{equation}
d_{\min}^{\rm ECAL}
=
\min_{i<j}
\left|
\mathbf{x}_{i}^{\rm ECAL}
-
\mathbf{x}_{j}^{\rm ECAL}
\right|,
\end{equation}
where $\mathbf{x}_{i}^{\rm ECAL}$ denotes the impact position of the
$i$th photon on the active ECAL surface.  This observable is more
conservative than considering only the two photon pairs originating
from $a_1$ and $a_2$, since a merged cluster can occur whenever any pair
among the five photons reaches the calorimeter with a sufficiently small
separation.

\begin{figure}[b!]
\centering
\includegraphics[width=0.82\linewidth]
{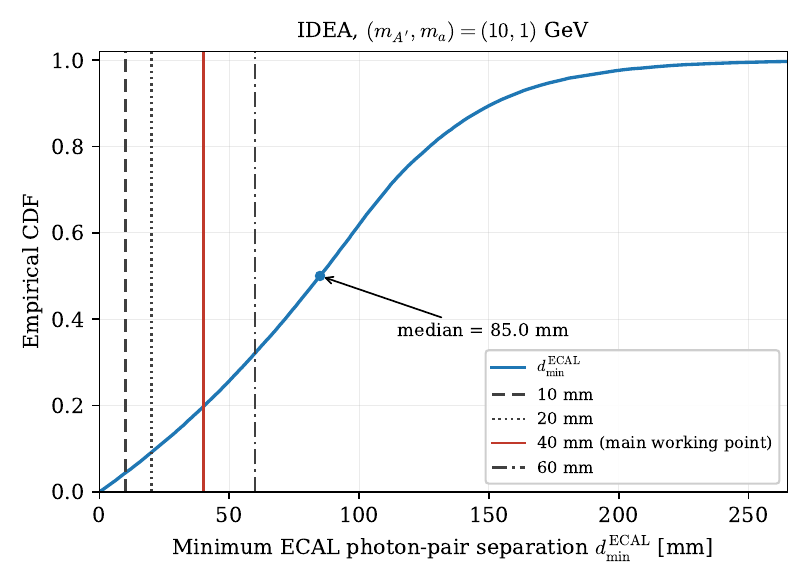}
\caption{
Empirical cumulative distribution of the minimum pairwise ECAL photon
separation, $d_{\min}^{\rm ECAL}$, for the benchmark point
$(m_{A'},m_a)=(10,1)\,\mathrm{GeV}$ in the IDEA detector geometry.
The distribution is evaluated for full-chain signal events satisfying
$E_{\gamma}^{\min}>2\,\mathrm{GeV}$, before applying the ECAL-separation
and non-pointing requirements.  The vertical lines mark separations of $10$, $20$, $40$, and $60\,\mathrm{mm}$; $40\,\mathrm{mm}$ is the main working point. The median minimum separation is approximately $85.0\,\mathrm{mm}$.
}
\label{fig:ecal-min-separation}
\end{figure}

Figure~\ref{fig:ecal-min-separation} shows the empirical cumulative
distribution of $d_{\min}^{\rm ECAL}$ for the benchmark point.  The
distribution is evaluated for full-chain signal events satisfying
$E_{\gamma}^{\min}>2~\mathrm{GeV}$ before the ECAL-separation and
non-pointing requirements are applied.  The median minimum separation is
\begin{equation}
{\rm median}\!\left(d_{\min}^{\rm ECAL}\right)
\simeq 85.0~\mathrm{mm},
\end{equation}
while the 10th percentile is approximately $21.8~\mathrm{mm}$.  We adopt $d_{\min}^{\rm ECAL}>40~\mathrm{mm}$ as the main working
point and use $20$ and $60~\mathrm{mm}$ as variations. The $40~\mathrm{mm}$
threshold corresponds to four nominal IDEA crystal-face widths and about
two PbWO$_4$ Moli\`ere radii~\cite{IDEADetector}. It is an effective
truth-level resolvability requirement, not a detector-validated
two-photon efficiency turn-on. The longitudinally
segmented front section may improve the separation of neighboring
showers by sampling them before maximum lateral development, but the
actual efficiency will depend on the two photon energies, incidence
angles, gaps, longitudinal information, and the clustering algorithm.

The same conditional sample also provides a direct separation-scale
sensitivity check. The fractions satisfying
$d_{\min}^{\rm ECAL}>10~\mathrm{mm}$ and
$d_{\min}^{\rm ECAL}>20~\mathrm{mm}$ are $0.956$ and $0.909$, respectively. Thus, changing the effective threshold from $20$ to $10~\mathrm{mm}$ increases the separation-only conditional retention by $0.048$ in absolute terms, or $5.2\%$ relative to the $20~\mathrm{mm}$ value. For $40$ and $60~\mathrm{mm}$ the fractions are about $0.80$ and $0.68$. This comparison is deliberately quoted only
for the distribution defined above; it is not a recalculation of the
total five-cluster acceptance.

The cumulative distribution in Fig.~\ref{fig:ecal-min-separation}
should not be interpreted as the total five-cluster reconstruction
efficiency.  The latter also includes the active IDEA ECAL acceptance,
the endcap beam-hole geometry, the photon-energy requirement, and the
condition that all five photons form distinct calorimeter clusters.  In
the $2.0\times10^{5}$ benchmark decay trials, requiring five resolved
ECAL clusters with a minimum pairwise separation of $20~\mathrm{mm}$ and
$E_{\gamma}>1~\mathrm{GeV}$ gives
\begin{equation}
\mathcal{A}_{5\gamma}
=
\frac{92982}{200000} = 0.465.
\end{equation}
After tightening the event-level photon-energy requirement to
$E_{\gamma}^{\min}>2~\mathrm{GeV}$, while retaining the five-cluster
condition, the acceptance becomes
\begin{equation}
\mathcal{A}_{5\gamma}^{E_{\min}>2~{\rm GeV}}
=
\frac{70297}{200000} = 0.351.
\end{equation}
These values distinguish the conditional geometrical separation
distribution from the full detector-oriented five-cluster acceptance.
At the $40~\mathrm{mm}$ working point the two acceptances are $0.414$ and
$0.311$. The selections that exploit the displaced decays are described
next.

\subsubsection{Direction-Free Selection with a Displaced-Vertex Fit}
\label{subsubsec:dv-selection}

A selection that relies on non-pointing photons needs the direction of
each electromagnetic shower. The calorimeter measures the shower
position precisely, but no FCC-ee study yet quantifies the
shower-direction resolution, and with two longitudinal crystal
segments it is likely to be much coarser than the position
resolution~\cite{Polesello2025}. We therefore first construct a
selection that uses only energies and impact positions. Non-pointing
information is treated in Sec.~\ref{subsubsec:nonpointing} as an
additional handle.

Under the hypothesis that all photons come from the interaction point,
the direction of photon $i$ is $\hat n_i=\vec X_i/|\vec X_i|$, where
$\vec X_i$ is the measured impact point, and its momentum is
$E_i\hat n_i$. Displaced decays might be expected to spoil the momentum balance or the reconstructed five-photon mass under this hypothesis. In practice, neither observable provides useful separation. Both ALPs
are produced at, or within a fraction of a millimeter of, the
interaction point, so each displaced vertex lies on the flight line of
its parent. The reconstructed direction of a photon pair then
coincides with the parent direction up to second-order terms, and the
pair momentum is wrong only by $(m_a^2-m_{\gamma\gamma}^2)/(2E)$. In the
benchmark sample without detector smearing, the transverse momentum
imbalance has a median of $4~\mathrm{MeV}$ and stays below
$43~\mathrm{MeV}$ in $99.9\%$ of the events, and $m_{5\gamma}$ agrees
with $\sqrt{s}$ to $10^{-5}~\mathrm{GeV}$. Both are far below the
resolution, so these quantities do not separate displaced signal from
prompt background.

The displacement shows up instead in the internal structure of the
pairs. The opening angle seen from the interaction point is smaller
than the true one, and the reconstructed pair mass is reduced roughly
as $m_a(1-d/r)$, where $d$ is the decay distance and $r$ the distance
to the calorimeter (see also Fig.~4 of Ref.~\cite{Polesello2025}). In
the benchmark sample the median pair mass is $0.73~\mathrm{GeV}$
instead of $1~\mathrm{GeV}$. The two pairs of an event are compressed by
different amounts, so the hypothesis of two pairs of equal mass fails
for the signal itself. The three-photon mass of $\gamma_{A'}$ and the
$a_2$ pair is also shifted, down to $9.79~\mathrm{GeV}$ for
$m_{A'}=10~\mathrm{GeV}$, because the pair momentum is
mis-reconstructed.

These shifts can be corrected without using shower-direction information. For a
photon pair with impact points $\vec X_1$, $\vec X_2$ and energies
$E_1$, $E_2$, we place the decay vertex on the reconstructed pair
direction $\hat n_P$ at the distance $d$ for which the pair mass equals
the hypothesized ALP mass,
\begin{equation}
2E_1E_2\left(1-\hat u_1\cdot\hat u_2\right)=m_a^2,
\qquad
\hat u_i=\frac{\vec X_i-d\,\hat n_P}{|\vec X_i-d\,\hat n_P|},
\label{eq:dv-fit}
\end{equation}
which has a solution $d>0$ whenever the pair mass under the
interaction-point hypothesis is below $m_a$. The corrected photon
directions $\hat u_i$ are then used to compute the three-photon mass
$m_T$. In the benchmark sample without smearing, this displaced-vertex correction returns $m_T$ to the interval
$9.978$--$10.020~\mathrm{GeV}$ for $99.8\%$ of the events.
Iterating the pair direction does not improve the result; we use the
first solution.

For each event we test all 30 assignments of the five photons to the
roles $(\gamma_{A'},a_1\to\gamma\gamma,a_2\to\gamma\gamma)$. With the
per-event mass resolutions $\sigma_T$ and $\sigma_P$ propagated from
Eq.~(\ref{eq:lucchini-resolution}), we define
\begin{equation}
\chi^2_{\rm DV}=\left(\frac{m_T^{\rm DV}-m_{A'}}{\sigma_T}\right)^2
+\sum_{P=1,2}\left[\max\!\left(0,\frac{m_P-m_a}{\sigma_P}\right)\right]^2,
\label{eq:chi2-dv}
\end{equation}
where only an excess of the pair mass $m_P$ over $m_a$ is penalized,
since displacement can only reduce it. The assignment with the smallest
$\chi^2_{\rm DV}$ is kept. The direction-free displaced-vertex (DV) selection then requires $|m_T^{\rm DV}-m_{A'}|<c_T\,\sigma_T$, $(m_P-m_a)/\sigma_P<c_P$ for both
pairs, a minimum energy sharing $\min(E_1,E_2)/(E_1+E_2)>0.05$ in each
pair, and $|\cos\theta_T|<0.90$ for the direction of the three-photon
system. Scalar ALP decays are isotropic in the ALP frame, whereas pairs
of QED photons are often strongly asymmetric in energy and the QED
background is peaked in the forward direction. The selection uses only
invariant masses and quantities that are insensitive to a longitudinal
boost; with leading-logarithmic ISR and the beam-energy spread
switched on, the signal efficiency of each requirement changes by
less than $0.5\%$. By contrast, cuts on the longitudinal momentum
balance or on $m_{5\gamma}-\sqrt{s}$ would change it by $12$--$16\%$,
and we do not use them. The cut values were chosen with a two-fold
cross-validation: they were optimized on half of the signal and
background samples and evaluated on the other half, and vice versa.
The two halves give $c_T=2.0$ or $2.5$ and $c_P=1.5$ or $1.0$. The yields
quoted below are the averages over the two test halves. The selection
uses the benchmark masses, so the quoted significances are local; a
search over a grid of mass hypotheses would require a trials
correction.

The background after this selection is far too small to be estimated
with unbiased simulation. With the photon-energy requirement
$E_\gamma>1~\mathrm{GeV}$ and $d_{\min}^{\rm ECAL}>40~\mathrm{mm}$ (see
Sec.~\ref{subsubsec:ecal-separation}), about $1.2\times10^{6}$ prompt five-photon
events pass the preselection at four interaction points. An unbiased
sample of practical size would contain only a few times $10^4$ such
events, none of which survive. We therefore generated a dedicated
sample with a generator-level requirement of two disjoint photon pairs,
each with invariant mass below $1.5~\mathrm{GeV}$. This requirement keeps
about $1.1\%$ of the cross section at each energy. It is looser than
the selection: every event in the unbiased samples that passes the
pair-mass requirement also passes the generator cut, and the largest
true mass of a selected pair is $1.07~\mathrm{GeV}$. The enriched samples
contain $2.02\times10^{5}$ events. As a closure test, they predict
$26.0$ events with two such pairs in the unbiased samples,
where $26$ are observed. After the full selection, $8$, $1$, and $0$ events
survive at $\sqrt{s}=91.19$, $87.9$, and $94.3~\mathrm{GeV}$,
respectively. Since fewer than ten events survive at each energy, we
quote the background as the sum of the per-energy $95\%$ confidence-level
upper limits,
\begin{equation}
N_B\le2.76\quad\text{(four interaction points)},
\label{eq:dv-background}
\end{equation}
while the simulated central value is $0.62$ events.

The signal and background yields are listed in Table~\ref{tab:dv-yields}, together with exact-Poisson thresholds. The discovery threshold $s_{5\sigma}$ is the signal for which the probability to observe at least $n_{\rm crit}$ events is $50\%$, where $n_{\rm crit}$ is the smallest count whose background-only probability is below $2.87\times10^{-7}$. The exclusion threshold $s_{95}$ is the median expected $95\%$ CL$_s$ upper limit, with ${\rm CL}_s(s)=P(n\le n_{\rm med}\,|\,s+b)/P(n\le n_{\rm med}\,|\,b)$ and $n_{\rm med}$ the median background-only count. The background upper limit of Eq.~(\ref{eq:dv-background}) is an ordinary one-sided Poisson confidence limit on the simulated counts.
For four interaction points, $576$ signal events are selected, and a
$5\sigma$ discovery requires $s_{5\sigma}=11.9$ events at the upper
limit of the background. When the background is varied by
$\pm10\%$, the threshold rises to at most $12.9$. The benchmark therefore lies well above the $5\sigma$ discovery threshold without any non-pointing information.

\begin{figure*}[t]
\centering
\includegraphics[width=0.96\textwidth]{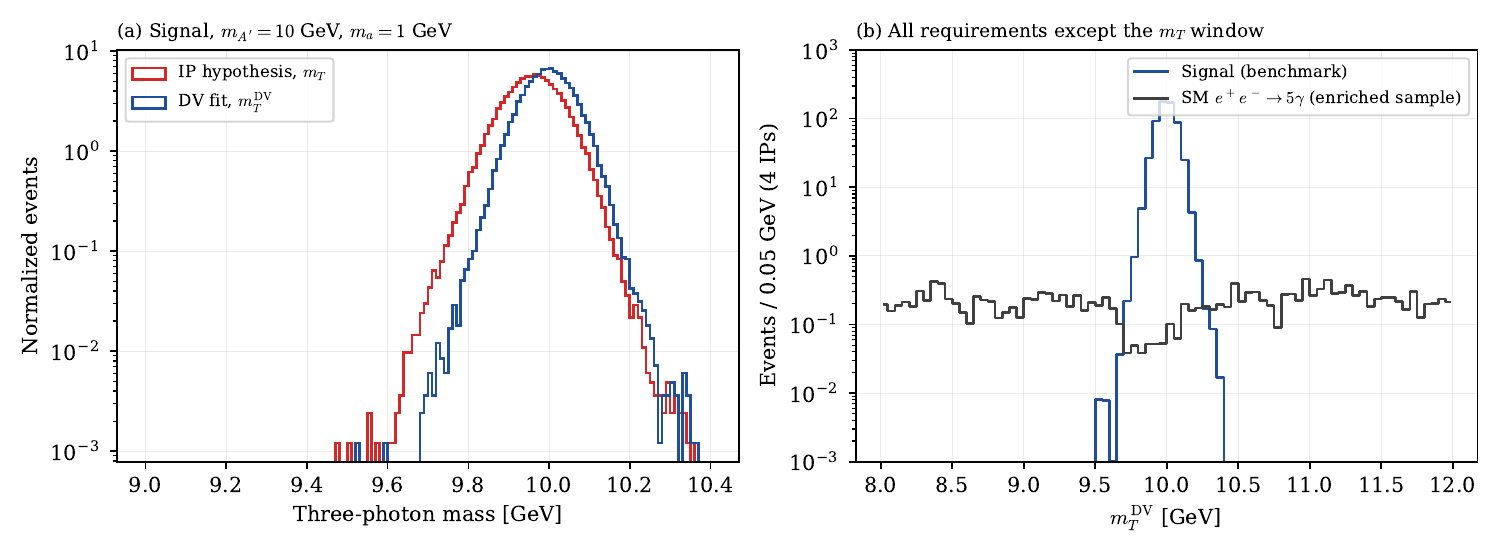}
\caption{(a) Three-photon mass of the benchmark signal reconstructed under
the interaction-point hypothesis ($m_T$) and after the displaced-vertex
correction ($m_T^{\rm DV}$), including the energy and position
resolution of Eq.~(\ref{eq:lucchini-resolution}). (b) $m_T^{\rm DV}$ after
all requirements of the DV selection except the $m_T$ window (shown here
for $c_P=1.5$), for the benchmark signal and for the prompt
$e^+e^-\to5\gamma$ background from the enriched simulation, normalized to
four interaction points.}
\label{fig:dv-mass}
\end{figure*}

Figure~\ref{fig:dv-mass} illustrates the effect of the fit for the
benchmark signal and for the prompt background.

\begin{table*}[t]
\centering
\caption{Selected signal and background events at the benchmark point with the DV selection ($E_\gamma>1~\mathrm{GeV}$,
$d_{\min}^{\rm ECAL}>40~\mathrm{mm}$), and, for comparison, with the
non-pointing selection of Sec.~\ref{subsubsec:nonpointing}
($E_\gamma>2~\mathrm{GeV}$, $d_{\min}^{\rm ECAL}>40~\mathrm{mm}$,
$\theta_{\rm NP}^{\max}>8~\mathrm{mrad}$, $p_{\rm tail}=10^{-6}$). The DV
background is the sum of per-energy $95\%$ CL upper limits from the enriched
simulation. $n_{\rm crit}$ is the smallest count giving a $5\sigma$ excess;
$s_{5\sigma}$ and $s_{95}$ are the median signal yields needed for a $5\sigma$ discovery and for a $95\%$ CL$_s$ exclusion. Couplings are in
GeV$^{-1}$; the $g_{a\gamma\gamma}$ interval has no upper edge within the
scanned range (see text).}
\label{tab:dv-yields}
\begin{ruledtabular}
\begin{tabular}{lccc}
 & DV, one IP & DV, four IPs & Non-pointing, four IPs\\
\hline
$N_S$ & $144.0$ & $576.1$ & $469.8$\\
$N_B$ & $\le0.69$ & $\le2.76$ & $2.79$\\
$n_{\rm crit}$ & $9$ & $15$ & $15$\\
$s_{5\sigma}$ & $7.98$ & $11.91$ & $11.88$\\
$s_{95}$ & $3.00$ & $5.52$ & $5.51$\\
$g_{a\gamma\gamma}$, $5\sigma$ & $>3.25\times10^{-7}$ & $>2.45\times10^{-7}$ & $(2.64$--$47.7)\times10^{-7}$\\
$g_{a\gamma\gamma'}$, $5\sigma$ & $>2.38\times10^{-7}$ & $>1.52\times10^{-7}$ & $>1.60\times10^{-7}$\\
$\epsilon$, $5\sigma$ & $<1.80\times10^{-5}$ & $<3.00\times10^{-5}$ & $<2.70\times10^{-5}$\\
\end{tabular}
\end{ruledtabular}
\end{table*}

\subsubsection{Non-Pointing Photons as an Additional Handle}
\label{subsubsec:nonpointing}

Where the shower direction can be measured, the angle
$\theta_{\rm NP}$ between the reconstructed photon direction and the
line from the interaction point to the impact position provides a
second, largely independent handle on the displacement. Non-pointing
photons are an established tool in LLP searches at hadron
colliders~\cite{ATLASNonPointing2013}. The non-pointing selection requires
$\theta_{\rm NP}^{\max}>8~\mathrm{mrad}$ for the most non-pointing photon.
Its prompt background is controlled by the tail of the angular response,
which we model as a per-photon probability $p_{\rm tail}$ of an
arbitrarily large mismeasurement on top of a Gaussian core whose width
scales as $\sigma_0\sqrt{5~\mathrm{GeV}/E_\gamma}$ below $5~\mathrm{GeV}$.
With the background normalization described above, the selection gives $469.8$ signal and $2.79$ background events for four interaction
points at $E_\gamma>2~\mathrm{GeV}$ and $p_{\rm tail}=10^{-6}$
(Table~\ref{tab:dv-yields}).

Since the shower-direction resolution is not known, it is more useful to
turn the argument around and ask what pointing performance a
non-pointing analysis would need. Scanning the core width $\sigma_0$ of a
two-dimensional Gaussian angular response, the tail probability, and the
cut, and requiring a $5\sigma$ discovery
with $N_B<1$ that survives a $\pm10\%$ change of the background, a
selection based on non-pointing alone needs
$\sigma_0\le8~\mathrm{mrad}$ and $p_{\rm tail}\le10^{-6}$ (with
$E_\gamma>5~\mathrm{GeV}$). When the non-pointing requirement is combined with the DV selection, the $N_B<1$ and $5\sigma$ criteria remain satisfied up to $\sigma_0=25.6~\mathrm{mrad}$, the largest value scanned, and for $p_{\rm tail}$ as large as $10^{-3}$, for each of the
photon-energy requirements $E_\gamma>1$, $2$, and $5~\mathrm{GeV}$.
Pointing information is therefore not required for the benchmark reach, but a modest pointing capability would reduce the simulated prompt $e^+e^-\to5\gamma$ background to a negligible level.

\paragraph{Trigger and online-selection assumption.}
The final FCC-ee trigger and data-acquisition architecture has not yet
been fixed: both a conventional triggered system and a trigger-less,
free-streaming scheme are under consideration, with the explicit design
goal of retaining annihilation events and unconventional LLP
signatures~\cite{FCCDetectorRequirements}. The signal contains five
resolved electromagnetic clusters, each with
$E_\gamma>1~\mathrm{GeV}$, and an event-scale electromagnetic energy of
order $m_Z$. These event characteristics are compatible with inclusive multi-cluster or total-ECAL-energy selections in a triggered architecture; in a free-streaming architecture, the same observables
could instead be formed during software-based online selection. We do
not emulate detector timing, front-end thresholds, online clustering,
or bandwidth limitations, so this expectation is not a detector-validated
trigger efficiency. No trigger factor is applied to the yields quoted
here, i.e., $\epsilon_{\rm trig}=1$.

\paragraph{Scope of the uncertainties.}
The quoted exact-Poisson thresholds are statistical reference values.
Signal and background expectations are treated as fixed, and no
nuisance parameters are profiled; for the DV selection the background
enters as an upper limit set by the size of the simulated samples. The
main effects that a future experimental likelihood would have to
include are: (i) the normalization of the leading-order
$e^+e^-\to5\gamma$ prediction and of the luminosity; (ii) reducible
backgrounds, for example $e^+e^-\to\gamma\gamma(\gamma)$ with additional
photons from detector effects or hadronic $Z$ decays with several
energetic $\pi^0$ mesons, which we have not simulated and which must be
studied through the DV selection; (iii) the photon energy and position
response, including non-Gaussian tails, the depth of the shower
centroid for inclined photons, and the separation of nearby showers;
(iv) the signal normalization and acceptance, including $N_Z$, trigger
efficiency, ISR, and the luminosity spectrum; and (v) for the
non-pointing selection, the rate and structure of the angular tails.
The ISR and beam-energy-spread morphing study above is likewise a
kinematic robustness test rather than a constraint on a nuisance
parameter.

\subsubsection{Coupling Dependence and Sensitivity}
\label{subsubsec:coupling-sensitivity}

\begin{figure}[t]
\centering
\includegraphics[width=0.98\linewidth]{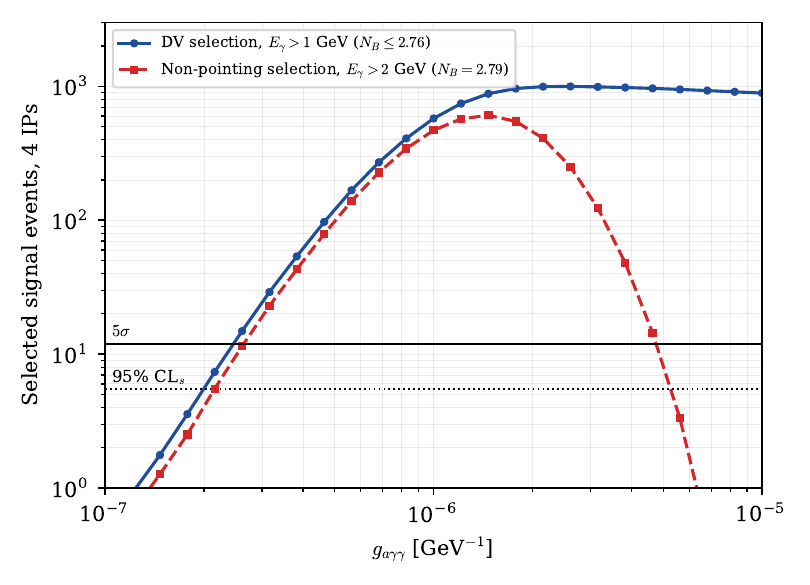}
\caption{Selected signal events for four interaction points as a function of
$g_{a\gamma\gamma}$, for $(m_{A'},m_a)=(10,1)~\mathrm{GeV}$,
$g_{a\gamma\gamma'}=10^{-6}~\mathrm{GeV}^{-1}$, and $\epsilon=10^{-7}$, with the
DV selection and with the non-pointing selection. The horizontal lines are the $5\sigma$ and $95\%$ CL$_s$ requirements for the DV selection; those of
the non-pointing selection agree within $0.3\%$.}
\label{fig:gagg-dv-vs-pointing}
\end{figure}

We next vary the couplings that control the cascade, with the masses
fixed to $(m_{A'},m_a)=(10,1)~\mathrm{GeV}$ and $\epsilon=10^{-7}$
unless stated otherwise, and apply the DV selection with the cuts
determined at the benchmark. The ALP decay lengths scale as
$g_{a\gamma\gamma}^{-2}$. Figure~\ref{fig:gagg-dv-vs-pointing} shows the
selected yield as a function of $g_{a\gamma\gamma}$ at fixed
$g_{a\gamma\gamma'}=10^{-6}~\mathrm{GeV}^{-1}$. At small coupling the ALPs
decay outside the calorimeter and both selections lose the signal in the
same way. The $5\sigma$ requirement is reached at
\begin{equation}
\begin{aligned}
g_{a\gamma\gamma}&\simeq2.45\times10^{-7}~\mathrm{GeV}^{-1}\quad(5\sigma),\\
g_{a\gamma\gamma}&\simeq2.00\times10^{-7}~\mathrm{GeV}^{-1}\quad(95\%~\mathrm{CL}_s),
\end{aligned}
\end{equation}
for four interaction points ($3.25\times10^{-7}$ and $2.46\times10^{-7}$ for one).
At large coupling the two selections differ qualitatively. The
non-pointing selection loses the signal as the decays become prompt, which
closes its sensitivity window at $4.77\times10^{-6}~\mathrm{GeV}^{-1}$. The DV
selection keeps it: the acceptance reaches a maximum of $0.566$ near
$g_{a\gamma\gamma}\simeq2.6\times10^{-6}~\mathrm{GeV}^{-1}$ and decreases only
slowly beyond, to $0.505$ at $10^{-5}~\mathrm{GeV}^{-1}$, because
undisplaced pairs populate both sides of the pair-mass requirement
whereas displaced pairs are always shifted below $m_a$. The selected
yield there is still about $891$ events, so the sensitivity extends
into the prompt regime and has no upper edge within the scanned range.

\begin{figure}[t]
\centering
\includegraphics[width=0.98\linewidth]{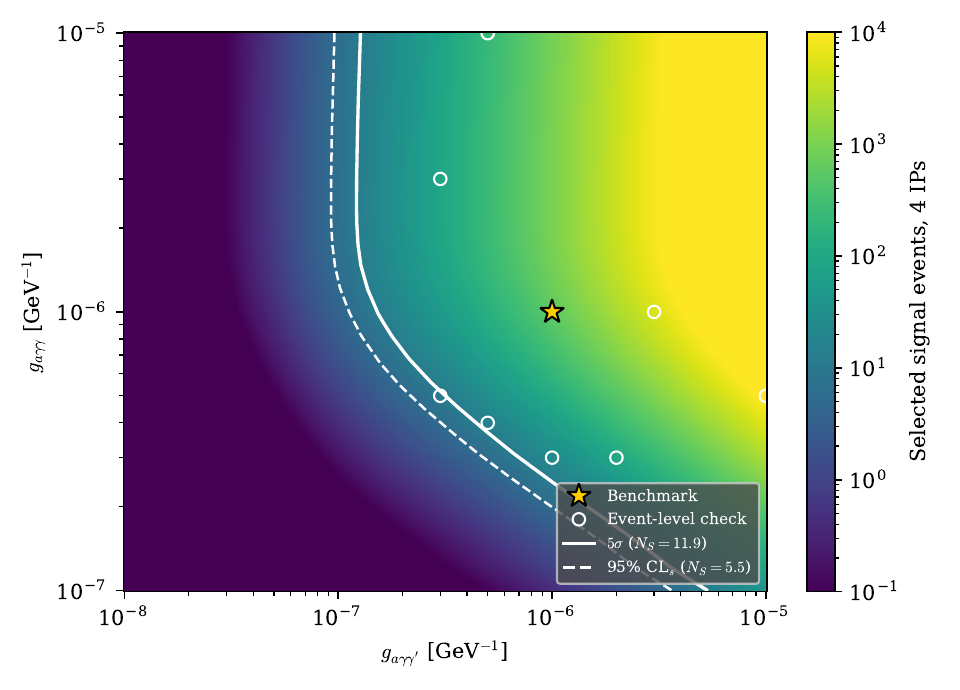}
\caption{Selected signal events for four interaction points in the
$(g_{a\gamma\gamma'},g_{a\gamma\gamma})$ plane for the DV selection, with
$(m_{A'},m_a)=(10,1)~\mathrm{GeV}$ and $\epsilon=10^{-7}$, from
Eq.~(\ref{eq:coupling-factorization}). The solid and dashed contours are the $5\sigma$ and $95\%$ CL$_s$ requirements. The star marks the benchmark and
the open circles the points where the factorization was checked against
event-level calculations.}
\label{fig:coupling-2d-dv}
\end{figure}

The coupling $g_{a\gamma\gamma'}$ controls both the production rate,
$\propto g_{a\gamma\gamma'}^2$, and the width of $A'\to a\gamma$. At fixed
$g_{a\gamma\gamma}=10^{-6}~\mathrm{GeV}^{-1}$ the dark photon decays well
inside the detector over the whole relevant range, so the lower edge is
set by the falling production rate, supplemented at the smallest
couplings by the decrease of $\mathrm{Br}(A'\to a\gamma)$. We find
\begin{equation}
\begin{aligned}
g_{a\gamma\gamma'}&\simeq1.52\times10^{-7}~\mathrm{GeV}^{-1}\quad(5\sigma),\\
g_{a\gamma\gamma'}&\simeq1.12\times10^{-7}~\mathrm{GeV}^{-1}\quad(95\%~\mathrm{CL}_s),
\end{aligned}
\end{equation}
for four interaction points ($2.38\times10^{-7}$ and $1.52\times10^{-7}$ for one). The
two-dimensional dependence is shown in Fig.~\ref{fig:coupling-2d-dv},
obtained from the factorized approximation
\begin{equation}
N_S(g_p,g_\gamma)\simeq
\frac{N_S(g_p,g_{\gamma,0})\,N_S(g_{p,0},g_\gamma)}{N_S(g_{p,0},g_{\gamma,0})},
\label{eq:coupling-factorization}
\end{equation}
with $g_{p,0}=g_{\gamma,0}=10^{-6}~\mathrm{GeV}^{-1}$. We checked it against
full event-level calculations at eight points spread over the plane; the
largest deviation is $1.2\%$ and the RMS deviation $0.8\%$.

The kinetic mixing $\epsilon$ enters mainly through
$\Gamma(A'\to f\bar f)\propto\epsilon^2$, which reduces
$\mathrm{Br}(A'\to a\gamma)$ at large $\epsilon$; the production rate and
the acceptance are nearly independent of it. With both portal couplings
at $10^{-6}~\mathrm{GeV}^{-1}$, the benchmark exceeds the $5\sigma$ threshold for
\begin{equation}
\epsilon\lesssim3.00\times10^{-5}\quad(5\sigma),
\end{equation}
for four interaction points ($1.80\times10^{-5}$ for one). The corresponding $95\%$ CL$_s$ reach, $4.43\times10^{-5}$, lies at the edge of the range of dark-photon
decay lengths covered by the simulation and is therefore an
extrapolation. These are detector-level reaches within the benchmark
and background model of this section and do not replace the external
dark-photon constraints of Sec.~IV.

\subsubsection{Mass Dependence}
\label{subsubsec:mass-dependence}

The mass scan uses the non-pointing selection of Sec.~\ref{subsubsec:nonpointing}
for one IP with $N_Z=1.5\times10^{12}$, $E_\gamma>2~\mathrm{GeV}$,
$d_{\min}^{\rm ECAL}>20~\mathrm{mm}$, and $\theta_{\rm NP}^{\max}>8~\mathrm{mrad}$.
The prompt background is normalized at a single energy,
$\sqrt{s}=m_Z$ with $37.5~\mathrm{ab}^{-1}$, using a sample generated with
$|\eta|<2.5$; it amounts to $N_B=0.44$ events, and the corresponding
median $5\sigma$ threshold is $s_{5\sigma}=7.23$. With the three-energy,
$|\eta|<3$ normalization of Sec.~\ref{subsubsec:idea-geometry}, the same
selection has $N_B=0.70$ and $s_{5\sigma}=7.97$. We use this scan to
illustrate how the ALP lifetime shapes the acceptance across the mass
plane. We do not repeat it with the DV selection because the enriched
background sample is restricted to pair masses below $1.5~\mathrm{GeV}$;
the heavier ALPs in the grid would require dedicated background samples.

The portal couplings and kinetic mixing are fixed to their benchmark values,
\begin{equation}
g_{a\gamma\gamma'}=g_{a\gamma\gamma}=10^{-6}~\mathrm{GeV}^{-1},
\qquad
\epsilon=10^{-7},
\end{equation}
and the same selection is applied throughout the scan.  We
consider
\begin{equation}
m_{A'}=5,\;10,\;30,\;60~\mathrm{GeV},
\qquad
m_a=0.3,\;1,\;3~\mathrm{GeV},
\end{equation}
giving 12 representative mass points.  The benchmark
$(m_{A'},m_a)=(10,1)~\mathrm{GeV}$ is included as one of these points.

It is useful to organize the scan with the dimensionless ratio
$r=m_a/m_{A'}$, which ranges from $0.005$ to $0.60$ on this grid.  In
the dark-photon rest frame, this ratio controls both the recoil momentum
and, at fixed portal coupling, the phase-space factor of the secondary
decay,
\begin{equation}
\begin{aligned}
E_{\gamma}^{*}=p_{a_2}^{*}
&=\frac{m_{A'}}{2}\left(1-r^2\right),
\\
\Gamma(A'\to a\gamma)&\propto
m_{A'}^3\left(1-r^2\right)^3.
\end{aligned}
\label{eq:mass-scan-r-kinematics}
\end{equation}
The ratio $r$ therefore helps identify compressed $A'\to a\gamma$
kinematics, but it does not reduce the two-dimensional scan to a single
variable.  The production phase space for $Z\to aA'$ depends on both
masses separately, while at fixed $g_{a\gamma\gamma}$ the ALP proper
lifetime scales as $m_a^{-3}$.

The finite-sample uncertainty is evaluated at the level of the parent
MadGraph events.  Each generated hard event is reused for $R=20$
stochastic decay realizations, so these realizations are not treated as
$R$ independent Bernoulli trials.  If $I_{ir}$ is the indicator that
decay realization $r$ of parent event $i$ passes the mass-scan selection, we define
\begin{equation}
q_i=\frac{1}{R}\sum_{r=1}^{R} I_{ir},
\qquad
\widehat{\mathcal A}_{\rm sel}
=\frac{1}{N_{\rm MG}}\sum_{i=1}^{N_{\rm MG}}q_i,
\end{equation}
and estimate the clustered standard error as
\begin{equation}
\delta_{\rm MC}\widehat{\mathcal A}_{\rm sel}
=
\left[
\frac{1}{N_{\rm MG}(N_{\rm MG}-1)}
\sum_{i=1}^{N_{\rm MG}}
\left(q_i-\widehat{\mathcal A}_{\rm sel}\right)^2
\right]^{1/2}.
\label{eq:mass-scan-clustered-mc-error}
\end{equation}
Confidence intervals are obtained from $10^5$ nonparametric bootstrap
replicas in which the parent MadGraph events, rather than the individual
decay realizations, are resampled.  The resulting signal-yield
uncertainties are obtained by applying the same bootstrap factors to the
fixed production normalization.  The displayed uncertainties therefore
quantify the finite detector-MC statistics.  The MadGraph integration
errors are sub-percent and are numerically subdominant.

The mass dependence of the selected signal is governed by three linked
ingredients.  First, increasing $m_{A'}$ reduces the available
$Z\to aA'$ phase space and therefore lowers the number of produced
signal cascades, most visibly at $m_{A'}=60~\mathrm{GeV}$.  Second, the
ALP mass strongly changes the decay length through the
$m_a^3$ dependence of
$\Gamma(a\to\gamma\gamma)$.  Third, $r$ controls the recoil in
$A'\to a\gamma$ through Eq.~(\ref{eq:mass-scan-r-kinematics}), affecting
the secondary-ALP boost and the energy of the accompanying photon.  As
a result, the detector acceptance is
largest only when the ALP decay length is comparable to the detector
dimensions: very light ALPs tend to decay outside the detector,
whereas heavier ALPs can become too prompt to satisfy the non-pointing
selection.

\begin{figure*}[t]
\centering
\includegraphics[width=0.98\textwidth]{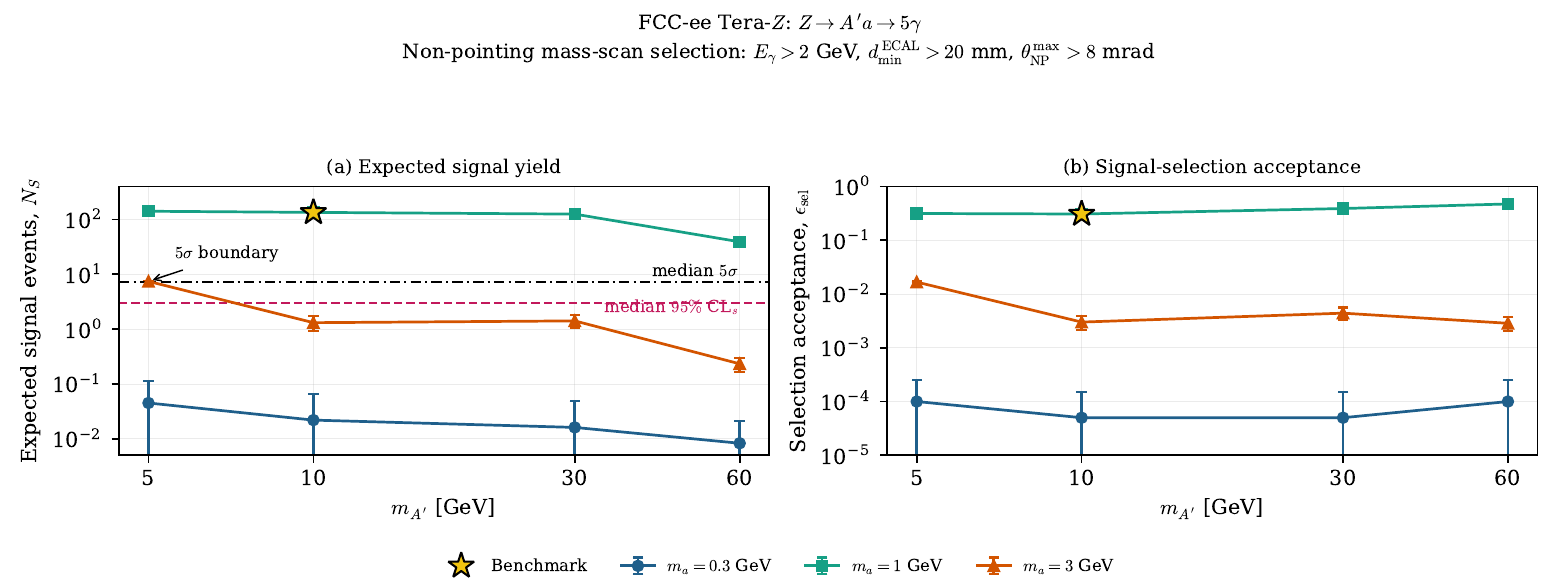}
\caption{
Mass dependence of the detector-level signal for the non-pointing mass-scan selection,
$E_{\gamma}^{\min}>2\,\mathrm{GeV}$,
$d_{\min}^{\rm ECAL}>20\,\mathrm{mm}$, and
$\theta_{\rm NP}^{\max}>8\,\mathrm{mrad}$, with
$g_{a\gamma\gamma'}=g_{a\gamma\gamma}=10^{-6}\,\mathrm{GeV}^{-1}$ and
$\epsilon=10^{-7}$.
The grid spans $0.005\leq r=m_a/m_{A'}\leq0.60$; because production
phase space and the ALP lifetime depend separately on the two masses,
the plotted curves are not functions of $r$ alone.
Panel (a) shows the expected selected signal yield for one FCC-ee
detector together with the median $5\sigma$ and $95\%$ CL$_s$
reference thresholds.  Panel (b) shows the corresponding total
selection acceptance.  Vertical error bars show the central $95\%$
parent-event bootstrap intervals; intervals whose lower endpoint is
zero extend below the displayed logarithmic range.  The star marks the
benchmark point
$(m_{A'},m_a)=(10,1)\,\mathrm{GeV}$.
The very small acceptance at $m_a=0.3\,\mathrm{GeV}$ is caused by ALPs
decaying predominantly outside the detector, whereas for
$m_a=3\,\mathrm{GeV}$ the loss is driven mainly by the increasingly
prompt ALP decays and the non-pointing requirement.
}
\label{fig:mass-dependence-mainwp}
\end{figure*}

Figure~\ref{fig:mass-dependence-mainwp} shows the expected selected
signal yield and the total main-working-point acceptance across the
scan.  For $m_a=0.3~\mathrm{GeV}$, the mean decay length of the directly
produced ALP lies between about $127$ and $223~\mathrm{m}$ over the
four dark-photon masses.  The probability for the complete cascade to
occur inside the detector is therefore extremely small, and the total
selection acceptance remains at the level of
$5\times10^{-5}$--$10^{-4}$.  The corresponding expected signal yield
never exceeds $4.5\times10^{-2}$ events.  This region is thus limited
primarily by the ALP lifetime rather than by photon reconstruction.

For $m_a=1~\mathrm{GeV}$, the mean ALP decay length instead falls in the
range
\begin{equation}
1.03~\mathrm{m}
\lesssim
\langle L_{a_1}\rangle
\lesssim
1.80~\mathrm{m},
\end{equation}
which is well matched to the IDEA detector dimensions.  The total
selection acceptance varies from approximately $0.31$ at
$m_{A'}=5$--$10~\mathrm{GeV}$ to $0.47$ at
$m_{A'}=60~\mathrm{GeV}$.  Nevertheless, the selected event yield
decreases at the largest dark-photon mass because of the reduced
production phase space.  The expected yields are approximately
$141.1$, $134.5$, $125.1$, and $39.1$ events for
$m_{A'}=5$, $10$, $30$, and $60~\mathrm{GeV}$, respectively.  All four
points remain above the median $5\sigma$ reference threshold even at the
lower endpoints of their clustered $95\%$ MC intervals for the adopted
background model.

The opposite behavior is seen for $m_a=3~\mathrm{GeV}$.  In this case,
the mean ALP decay length is only
$1.3$--$2.2~\mathrm{cm}$, so essentially all cascades occur inside the
detector and the five-cluster probability is high.  The dominant loss
instead arises from the non-pointing requirement because the ALP
decays are now close to the interaction point.  The resulting total
selection acceptance is at the $10^{-3}$--$10^{-2}$ level.  At the
$(m_{A'},m_a)=(5,3)~\mathrm{GeV}$ point, $r=0.60$ and
$E_{\gamma}^{*}=1.60~\mathrm{GeV}$, so the compressed secondary decay
also makes the photon-energy requirement less efficient.  Nevertheless,
the dominant loss is still non-pointing: $0.675$ of the decay trials
pass the main requirements before that cut, whereas its conditional
efficiency is only $0.0247$.  The comparatively large preselection
normalization of about $446$ produced cascades at low $m_{A'}$ leaves
the point close to the discovery threshold despite a total selection
acceptance of only $0.0167$.  Its selected yield is
\begin{equation}
N_S=7.43\pm0.13\quad(1\sigma~\mathrm{MC}),
\qquad
N_S^{95\%\,\mathrm{MC}}\in[7.18,\,7.67].
\end{equation}
Although the central value lies slightly above the median $5\sigma$
reference value $s_{5\sigma}=7.23$, the clustered $95\%$ MC interval
crosses this threshold.  Equivalently, $94.5\%$ of the bootstrap
replicas lie above the median $5\sigma$ requirement.  We therefore classify this mass point as a boundary case rather than a robust $5\sigma$ point. With the three-energy normalization ($s_{5\sigma}=7.97$) it lies below the threshold.  The other $m_a=3~\mathrm{GeV}$ points fall below the median
$5\sigma$ requirement.

The numerical values are collected in
Table~\ref{tab:mass_dependence_mainWP}.  Most off-benchmark points were
evaluated with $10^3$ generated MadGraph events and 20 stochastic decay
realizations per event.  The benchmark uncertainty is evaluated with
the corresponding $10^4$-event validation sample.  Because the
$(5,3)~\mathrm{GeV}$ point lies
close to the median $5\sigma$ boundary, it was re-evaluated with a
higher-statistics sample corresponding to $2.1\times10^4$ generated
events and $4.2\times10^5$ decay trials.  The table quotes the clustered
one-standard-error uncertainties, whereas
Fig.~\ref{fig:mass-dependence-mainwp} displays the corresponding
clustered $95\%$ bootstrap intervals.

\begin{table*}[t]
\centering
\caption{Mass dependence of the detector-level signal with the non-pointing mass-scan selection for one FCC-ee detector. The benchmark point $(m_{A'},m_a)=(10,1)\,\mathrm{GeV}$ is shown in bold. The last column refers to the median $5\sigma$ threshold $s_{5\sigma}=7.23$ of this scan (see text).}
\label{tab:mass_dependence_mainWP}
\begin{ruledtabular}
\begin{tabular}{cccccc}
$m_{A'}$ [GeV] & $m_a$ [GeV] & $\langle L_{a_1}\rangle$ [m] & $\mathcal{A}_{\mathrm{sel}}$ & $N_S$ & Median $5\sigma$ \\
$5$ & $0.3$ & $222.651$ & $1.00\times 10^{-4}$ & $0.0450$ & No \\
$5$ & $1$ & $1.803$ & $0.31385$ & $141.1$ & Yes \\
$5$ & $3$ & $2.22\times 10^{-2}$ & $0.01665$ & $7.426$ & Yes \\
$10$ & $0.3$ & $220.637$ & $5.00\times 10^{-5}$ & $0.0219$ & No \\
$\mathbf{10}$ & $\mathbf{1}$ & $\mathbf{1.787}$ & $\mathbf{0.30705}$ & $\mathbf{134.5}$ & \textbf{Yes} \\
$10$ & $3$ & $2.20\times 10^{-2}$ & $0.003$ & $1.310$ & No \\
$30$ & $0.3$ & $199.150$ & $5.00\times 10^{-5}$ & $0.0161$ & No \\
$30$ & $1$ & $1.613$ & $0.38825$ & $125.1$ & Yes \\
$30$ & $3$ & $1.99\times 10^{-2}$ & $0.0044$ & $1.412$ & No \\
$60$ & $0.3$ & $126.632$ & $1.00\times 10^{-4}$ & $8.29\times 10^{-3}$ & No \\
$60$ & $1$ & $1.025$ & $0.47215$ & $39.09$ & Yes \\
$60$ & $3$ & $1.26\times 10^{-2}$ & $0.00285$ & $0.2329$ & No \\
\end{tabular}
\end{ruledtabular}
\end{table*}

With the non-pointing selection the scan shows a finite
lifetime window.  Sensitivity is reduced when the ALP
is so long lived that it escapes the detector and also when it is so
short lived that the photons appear effectively prompt.  The
$m_a\simeq1~\mathrm{GeV}$ points considered here lie between these two
regimes and retain sizeable detector-level acceptance over a broad
range of dark-photon masses.
The upper side of this window is a property of the non-pointing
selection; the DV selection of Sec.~\ref{subsubsec:dv-selection} retains
prompt decays (Fig.~\ref{fig:gagg-dv-vs-pointing}).

\section{Conclusion}
\label{sec:conclusion}

We have studied a massive dark photon and an axion-like particle
connected through a gauge-invariant mixed portal at the FCC-ee $Z$ pole.
For the dimension-five hypercharge portal the mixed couplings satisfy
$g_{aZ\gamma'}=-\tan\theta_W\,g_{a\gamma\gamma'}$, as first noted in
Ref.~\cite{Jodlowski2025}, so the interaction that allows
$A'\to a\gamma$ also produces $Z\to aA'$. When both ALPs decay to
photons, this gives the fully visible cascade
\begin{equation}
Z\to a_1A'\to a_1a_2\gamma\to5\gamma.
\end{equation}

We derived the decay widths, branching fractions, two-body kinematics,
and laboratory decay lengths. The dark-photon width into Standard Model
states includes the admixture of the $Z$ boson, which is negligible at
the benchmark but increases the width by a factor of up to $13$ near the
kinematic endpoint and changes the low-coupling behavior of the
branching fraction there. Because the secondary ALP starts at the
dark-photon decay vertex, the fiducial probability requires a sequential
treatment of the decay positions. For
$m_{A'}=10~\mathrm{GeV}$, $m_a=1~\mathrm{GeV}$, $\epsilon=10^{-7}$, and
$g_{a\gamma\gamma'}=g_{a\gamma\gamma}=10^{-6}~\mathrm{GeV}^{-1}$, we obtain
$\mathrm{Br}(Z\to aA')=2.92\times10^{-10}$ and
$\mathrm{Br}(A'\to a\gamma)=0.9995$, i.e., about $438$ cascades for
$1.5\times10^{12}$ $Z$ bosons and about $440$ per interaction point over
the three-energy program.

At the benchmark, $41\%$ of the cascades satisfy the five-cluster requirement with $d_{\min}^{\rm ECAL}>40~\mathrm{mm}$ and $E_\gamma>1~\mathrm{GeV}$. The detector-level analysis shows how the displacement can be reconstructed without shower-direction information.
Under the hypothesis that all photons come from the interaction point,
the momentum balance and the five-photon mass of a displaced signal
event are preserved to within tens of MeV, because each ALP decays on
the flight line of its parent. The displacement instead compresses the reconstructed diphoton masses. A displaced-vertex (DV) fit, which places the decay vertex of each pair along its reconstructed direction so that the pair mass equals $m_a$, restores the three-photon mass of the $A'$ and the full resonance
structure of the cascade, using only photon energies and impact
positions. The DV selection built on this fit retains $576$ signal events for
four interaction points. The remaining prompt background is too small for
an unbiased simulation; a dedicated sample enriched in events with two
low-mass photon pairs, validated by a closure test, bounds it to at most $2.8$ events at $95\%$ CL. The benchmark therefore lies well above the $5\sigma$ discovery threshold without using shower-direction information.

For four interaction points, the $5\sigma$ reach in $g_{a\gamma\gamma}$ extends down to $2.5\times10^{-7}~\mathrm{GeV}^{-1}$, below which the ALPs escape the calorimeter. No upper edge appears within the scanned range because the DV selection retains prompt decays. In the mixed coupling the $5\sigma$ reach extends down to $g_{a\gamma\gamma'}\simeq1.5\times10^{-7}~\mathrm{GeV}^{-1}$, and the benchmark stays above the $5\sigma$ threshold for $\epsilon\lesssim3\times10^{-5}$.
Within the angular-response model used here, non-pointing photons alone would require a core shower-direction resolution of about $8~\mathrm{mrad}$ and a tail probability below $10^{-6}$ per photon to keep the background below one event; combined with the DV selection, this remains possible up to the largest core width scanned, $25.6~\mathrm{mrad}$, for tail probabilities up to $10^{-3}$. The mass scan with the non-pointing selection shows how the ALP
lifetime shapes the acceptance across the mass plane. Extending the DV selection to heavier ALPs requires dedicated background
samples.

In the same model, the ordinary coupling
$g_{aZ\gamma}=-\tan\theta_W\,g_{a\gamma\gamma}$ induces $Z\to\gamma a$. At
the benchmark its branching fraction is of the same order as that of
$Z\to aA'$, and $g_{a\gamma\gamma}=10^{-6}~\mathrm{GeV}^{-1}$ corresponds
to $C_{\gamma\gamma}/\Lambda\simeq2.5$--$2.7\times10^{-3}~\mathrm{TeV}^{-1}$
in the conventions of Ref.~\cite{Polesello2025}, close to the projected
three-photon reach. The three-photon channel would provide a correlated signal, while the five-photon cascade is directly sensitive to the mixed-portal parameters $m_{A'}$ and $g_{a\gamma\gamma'}$.

These central-detector sensitivities are complementary to the LLP
program proposed for the HL-LHC.  FASER2 targets light, very forward,
highly boosted states that traverse the long baseline downstream of the
ATLAS interaction point before decaying; MATHUSLA uses a much larger
surface decay volume to access longer lifetimes; and CODEX-b employs a
compact shielded volume near LHCb, with benchmark studies including LLP
production in heavy-flavor and Higgs decays
\cite{FASERPhysicsReach,Mathusla,Codexb}.  By contrast, the present
FCC-ee strategy uses resonant $Z$ production in a clean $e^+e^-$
environment and reconstructs an exclusive displaced five-photon cascade inside a meter-scale central detector.  These approaches thus
sample different production mechanisms, boost and angular regimes, and
proper-lifetime ranges.  We do not make a direct numerical comparison:
translating the projected FASER2, MATHUSLA, or CODEX-b reach to this
mixed dark-photon--ALP portal, particularly for a photon-only displaced
decay, would require dedicated production and detector-response recasts.

The present study is a fast detector analysis rather than a full
simulation. The calorimeter is described by its geometry and by
parametrized energy and position resolutions; shower development for
inclined photons, the separation of nearby showers, and the online
selection are not simulated, and the yields assume a trigger efficiency of one.

The background estimate is limited by the size of the enriched
sample and enters as an upper limit, the $e^+e^-\to5\gamma$ prediction is
leading order, and reducible backgrounds, such as $e^+e^-\to\gamma\gamma(\gamma)$
with additional photons from detector effects or hadronic $Z$ decays
with several energetic $\pi^0$ mesons, are not evaluated here and require a dedicated study with the DV selection. The significances are local to
the benchmark masses. The benchmark regions should be read together with
the external constraints of Sec.~IV rather than as a global recast.

Within these qualifications, the gauge-invariant dark-photon--ALP portal
produces a distinctive five-photon signature at the FCC-ee $Z$ pole. The
large $Z$ sample compensates for the small $\mathrm{Br}(Z\to aA')$, and
the resonance structure of the cascade, recovered with the
displaced-vertex fit, suppresses the simulated prompt $e^+e^-\to5\gamma$ background to the level of a few events without relying on calorimeter pointing.

\bibliography{references}

\end{document}